\documentclass[a4paper,11pt]{article}
\usepackage{jheppub} 
\pdfoutput=1 
\usepackage{graphicx,dcolumn,bm,epsfig,cancel,hyperref}
\usepackage{multirow}
\usepackage{amsmath}
\usepackage{amssymb, times}
\newcommand{\comment}[1]{{}}
\usepackage{color}
\newcommand{\mbs}[0]{{M_{\text{BS}}}}
\newcommand{\rbs}[0]{{R_{\text{90}}}}
\newcommand{\cbs}[0]{{C_{\text{BS}}}}
\newcommand{\rtidal}[0]{{r_{\text{Tidal}}}}

\title{
Probing Boson Stars with Extreme Mass Ratio Inspirals
}
\author[a]{Huai-Ke Guo}
\author[a]{Kuver Sinha}
\author[b, c]{Chen Sun}
\affiliation[a]{Department of Physics and Astronomy, University of Oklahoma, Norman, OK 73019, USA}
\affiliation[b]{CAS Key Laboratory of Theoretical Physics, Institute of Theoretical Physics, Chinese
Academy of Sciences, Beijing 100190, P. R. China}
\affiliation[c]{Department of Physics, Brown University, Providence, RI, 02912, USA}
\emailAdd{ghk@ou.edu}
\emailAdd{kuver.sinha@ou.edu}
\emailAdd{chen.sun@brown.edu}

\begin{document}


\abstract
{We propose to use gravitational waves from extreme mass ratio inspirals (EMRI), composed of a boson star and a supermassive black hole in the center of galaxies, as a new method to search for boson stars. Gravitational waves from EMRI have the advantage of being long-lasting within the frequency band of future space-based interferometer gravitational wave detectors and can accumulate large signal-to-noise ratio (SNR) for very sub-solar mass boson stars. Compared to gravitational waves from boson star binaries, which fall within the LIGO band, we find that  much larger ranges of the  mass and compactness of boson stars, as well as the underlying particle physics parameter space, can be probed by EMRI. We take tidal disruption of the boson stars into account and distinguish those which dissolve before the inner-most-stable-circular-orbit (ISCO) and those which dissolve after it. Due to the large number of cycles recorded, EMRIs can lead to a very precise mass determination of the boson star and distinguish it from standard astrophysical compact objects in the event of a signal. Tidal effects in inspiralling binary systems, as well as possible correlated electromagnetic signals, can also serve as potential discriminants. 
}


\maketitle

\section{Introduction}

Light scalars with long de Broglie wavelengths and suppressed
couplings to the Standard Model can form gravitationally bound
Bose-Einstein condensates (BEC) leading to macroscopic objects called
boson stars~\cite{Colpi:1986ye,Tkachev:1986tr,vanderBij:1987gi}. The study of such condensates has a rich history in the context of early Universe cosmology. Analytic and numerical investigations pertaining to the evolution, stability, and decay of boson stars have been undertaken by various
groups (we refer to~\cite{JetzerP,Schunck:2003kk,Liebling:2012fv} for reviews). These questions are also of interest to particle physicists, given the central place that light fundamental scalars occupy in theories beyond the Standard Model. The hope is that astrophysical constraints on boson stars will translate into constraints on the scalar  field itself, such as its mass and interactions.

Gravitational waves~\cite{Abbott:2016blz} can play a significant role in these questions. 
%
%
The general idea is that for
macroscopic exotic compact objects (ECOs) formed of light bosons, with
masses comparable to stellar black holes and compactness sufficiently
large that a binary system can form, gravitational wave signals will be generated. Current and future gravitational wave detectors can constrain the mass and compactness of the ECOs. Since these attributes are in turn determined by the physics at the microscopic scale such as the mass and self-interaction of the field, the probed regions of the mass-compactness plane will shed light on the underlying particle physics parameters.  Several cases have
been pursued by different groups: mergers of mini-boson and solitonic
boson stars \cite{1989PhRvL..63.1199G, Palenzuela:2006wp, Palenzuela:2007dm,Amin:2013eqa,
  Bezares:2018qwa}, oscillations \cite{Helfer:2018vtq}, Proca stars
\cite{Sanchis-Gual:2018oui}, and repulsive boson stars \cite{Croon:2018ybs,Croon:2018ftb}. These studies focused on binary systems
with comparable masses, similar to stellar black holes, which produce
gravitational waves within the band of LIGO~\cite{Giudice:2016zpa}.

The purpose of this paper is to point out that extreme mass ratio
inspirals (EMRI) are very effective in probing boson stars and future
space-based interferometers which target EMRI can potentially
constrain a large part of the mass versus compactness parameter
space. The standard EMRI consists of a stellar black hole, neutron
star or white dwarf circling around a supermassive black hole (SMBH)
at the center of each galaxy. ECOs, if they exist, can also create
EMRI systems and constitute an entirely new target species for
space-based gravitational wave detectors. Compared to the binary
system detected by LIGO, an ECO can linger around the
inner-most-stable-circular-orbit (ISCO) of the SMBH for a very long
time and accumulate a large signal-to-noise ratio (SNR) in the LISA
band even it has very sub-solar mass. Another factor further
enhances the SNR: since astrophysical observations suggest that the
spins of SMBHs in galaxies tend to have a value very close to 1, this
results in a decreased radius of the ISCO and allows the ECOs to be
closer to the SMBH, leading to larger SNR.

An important aspect of our work is to stress that in the event of a
discovery, EMRIs would naturally allow for an unambiguous
identification of the participant as an ECO and not a standard
astrophysical compact object. Due to the large number of cycles recorded, even for very
sub-solar ECOs, EMRIs can lead to a very precise determination of the
parameters in the system, such as the masses of the SMBH and the boson
star, and the SMBH spin, with errors as small as
$10^{-5}$~\cite{Barack:2003fp,Klein:2015hvg}. Therefore if gravitational waves are observed from an EMRI
system and it is determined that the smaller object has sub-solar mass,
this would definitely rule out the possibility of impostors like
stellar black holes, neutron stars or white dwarfs. In
  particular, for those ECOs that are disrupted before reaching ISCO,
  we can also estimate their radius (hence compactness) to break the
  degeneracy with primordial black holes. Moreover, tidal effects in inspiralling binary systems, as well as possible correlated electromagnetic signals, can also serve as a potential discriminant between ordinary compact objects and boson stars.

The typically large SNR for even very sub-solar mass ECOs thus provides an  extensive probe of the ECO parameter space, extending the coverage of mass and compactness to much lower regions, and serves as a powerful probe of the underlying particle physics.

Before proceeding, we note that in our work we remain agnostic about
the connection between dark matter and boson stars, an aspect that has
been explored by several authors. In particular, we do not impose dark
matter constraints on the particle physics model, except that we
require the scalar to be complex so that it has conserved particle
number, massive, and having repulsive quartic interaction. Our aim is
to present a case for using EMRIs to probe boson stars, and we study
this case using a phenomenological model. The drawback in such an
approach is that it is then difficult to provide the predicted EMRI
event rate, since abundances are free to vary. We briefly provide an
event rate estimation in Section~\ref{sec:results}, assuming that ECOs
constitute the entire dark matter halo at the center of the galaxy,
although we do not further explore the consequences of such an
assumption in our model, at the level of the particle physics.

The paper is organized as follows. We start with a description of
boson stars in Section~\ref{sec:bs}, where we discuss the
phenomenological model and compute the profile of the boson star in the
mass-compactness plane. We then discuss the EMRI system as well as
gravitational waves from such systems in Section~\ref{sec:EMRI}.
  In Section~\ref{sec:results}, we show the results of our analysis,
  including the region of boson star profile that LISA is sensitive
  to, the corresponding scalar theory parameter space, and an estimate
  of the event rate. We also briefly discuss possible connections to dark
  matter physics. In section~\ref{sec:summary}, we conclude with a summary
  of the main results.




\section{\label{sec:bs}Boson Stars}

Boson stars are macroscopic quantum states formed in the early
Universe, protected against gravitational collapse by the Heisenberg
Uncertainty Principle (for non-interacting or attractively
self-interacting bosons \cite{Ruffini:1969qy, Levkov:2016rkk,Croon:2018ybs}) or by
repulsive self-interactions \cite{Colpi:1986ye,Gleiser:1988ih,Croon:2018ybs}. If the star is
composed of real scalars, it can decay with lifetime shorter than the
age of the Universe \cite{Gleiser:2009ys}. In addition to the decay
led by the intrinsic dispersion of the wave packet, nonlinear mode
coupling that arises from self-interaction \cite{Eby:2015hyx,
  Eby:2017teq, Visinelli:2017ooc} and particle decay due to coupling
with other species \cite{Hertzberg:2018zte} constitute extra decay
channels of the boson star.\footnote{In \cite{Eby:2015hyx} it is
  argued that very small and dilute axion stars may have lifetimes comparable to the age of the Universe. Nevertheless, we do not consider the case of real scalars any further in this work.} 
%
%
In what follows, we
assume a phenomenological model of complex scalars which are naturally
long-lived since the particle number is protected by the underlying
$U(1)$ symmetry. We discuss the mass profile of the resulting boson
star.

We first note that boson stars are unstable against gravitational
collapse into black holes above a critical maximum mass against
central density \cite{Hawley:2000dt, Gleiser:1988ih}. In addition to
this critical bound, the mass is also constrained by requiring wave
function stability against radial perturbations, which arise from the
nonlinear dispersion. This stability bound is explicit when the bosons
are attractively self-interacting \cite{Eby:2015hyx, Braaten:2015eeu,
  Schiappacasse:2017ham, Eby:2017teq, Visinelli:2017ooc,
  Chavanis:2017loo}. On the other hand, the bound is implicit in the
case of non-interacting and repulsive theories and is revealed only
when spacetime backreaction is taken into account
\cite{Croon:2018ybs}. For negligible self-interactions, boson stars
have a maximal mass $\sim M_{Pl}^2/m$, where $m$ is the mass of the
scalar. On the other hand, for bosons with a repulsive $\lambda\phi^4$
interaction, the maximal mass scales as
$\sim \sqrt{\lambda}M_{Pl}^3/m^2$.  We refer to \cite{Schunck:2003kk}
for a review of boson stars resulting from different types of boson
self-interactions, and \cite{Fan:2016rda} for particle physics models
that lead to repulsive $\phi^4$ interaction. In what follows we focus
on boson stars that result from complex $|\phi|^4$ theory of repulsive
interaction, whose strength ranges from being stronger than gravity to being
negligible (essentially a $m^2 |\phi|^2$ theory.)

\subsection{The Particle Physics Model}
\label{sec:particle-model}

We first discuss the particle physics model and set our notation and conventions. Up to renormalizable terms, the complex scalar theory is given by
\begin{equation}
  \label{eq:Lagrangian-KG-curved-space}
  \mathcal{L}=    \frac{1}{2}g^{\mu \nu}\partial_\mu \phi^* \partial_\nu \phi -
  \frac{1}{2}m^2 |\phi|^2 - \frac{\lambda}{4}\left
    (\frac{m^2}{f^2} \right )|\phi|^4,
\end{equation}
where $f$ has mass dimension one, and $\lambda$ is of order one and positive.
Higher dimensional operators are suppressed by powers of $(m/f)$. This parametrization is motivated by identifying the scalar field as a pseudo Nambu-Goldstone boson (pNGB) whose mass is protected by an approximate shift symmetry. 
Interactions between $\phi$ and the Standard Model are  assumed to be suppressed and do not change the wave function significantly. We use the $(+,-,-,-)$ signature, and assume spherical symmetry. 
The energy momentum tensor is given by 
\begin{eqnarray}
  T_\mu^\nu
  & =&
        \frac{\delta \mathcal{L}}{\delta (\partial_\nu \phi)}
    \partial_\mu\phi
        +
    \frac{\delta \mathcal{L}}{\delta (\partial_\nu \phi^*)} \partial_\mu
    \phi^*
    -
     \delta^\nu_\mu \mathcal{L}    
    \cr
  & =&
       \frac{1}{2}     g^{\nu \nu'} \partial_{\nu'} \phi^* \partial_\mu
    \phi 
       + \frac{1}{2} g^{\nu\nu'} \partial_{\nu'} \phi \partial_\mu
       \phi^*
       -\delta_\mu^\nu \left ( \frac{1}{2} g^{\mu' \nu'}\partial_{\mu'} \phi^*
       \partial_{\nu'} \phi - \frac{1}{2}m^2 |\phi|^2 -
    \frac{\lambda}{4}\left
       (\frac{m^2}{f^2} \right )|\phi|^4 \right ).
       \cr
       &&
\end{eqnarray}

\subsection{The Einstein-Klein-Gordon System}
\label{sec:einst-klein-gord}

Depending on the compactness of the boson star, the metric can  deviate significantly from the flat limit. 
Assuming spherical symmetry, the metric can be parametrized as
\begin{equation}
  \label{eq:metric}
  ds^2 =
    B(r) dt^2 - A(r) dr^2 - r^2 d\theta^2 - r^2 \sin^2 \theta d
    \phi^2. 
\end{equation}
The Einstein tensor $G_\mu^\nu$ is diagonal, with the following non-zero
components:
\begin{eqnarray}
G_t^t& =&   -\frac{A'(r)}{r A(r)^2}+\frac{1}{r^2
       A(r)}-\frac{1}{r^2},
       \cr
       G_r^r
  & =& \frac{B'(r)}{r A(r)
   B(r)}+\frac{1}{r^2
               A(r)}-\frac{1}{r^2}.
\end{eqnarray}
In the above, $A(r), B(r),$ and $\phi(r)$ are three scalar degrees of freedom. Two constraints are obtained from solving  the ${}^t_t$ and ${}^r_r$ components of Einstein equation, while the third is obtained from the equations of motion of the scalar field. Together, they form the
Einstein-Klein-Gordon system:
\begin{align}
  \label{eq:KG-eom-1-2}
& \frac{4\pi G_N}{B(r)} \partial_t \phi \partial_t \phi^*
  +  \frac{4\pi G_N}{A(r)} \partial_r \phi \partial_r \phi^*
                          +4\pi G_N m^2 |\phi|^2 + 2G_N \pi \lambda
                          \left (\frac{ m^2}{ f^2}\right ) |\phi|^4 - \frac{A'(r)}{r
  A(r)^2} + \frac{1}{r^2 A(r)} - \frac{1}{r^2} = 0,
  \cr
& \frac{4\pi G_N}{B(r)} \partial_t \phi \partial_t \phi^*
  +  \frac{4\pi G_N}{A(r)} \partial_r \phi \partial_r \phi^*
        -4\pi G_N m^2 |\phi|^2 - 2G_N \pi \lambda
                          \left (\frac{ m^2}{ f^2}\right )
        |\phi|^4
-\frac{B'(r)}{r A(r) B(r)} - \frac{1}{r^2 A(r)} + \frac{1}{r^2}
        = 0,
        \cr
&  \frac{1}{A}\partial_r^2 \phi - \frac{1}{B}  \partial_t^2 \phi +
  \partial_r \phi \left (\frac{B'(r)}{2 A(r)B(r)} - \frac{A'(r)}{2A(r)^2} +
  \frac{2}{A(r)r}\right ) - m^2 \phi - \lambda\left (
  \frac{m^2}{f^2} \right ) |\phi|^2 \phi = 0.        
\end{align}
%
We assume the harmonic ansatz $\phi(r,t) = \Phi \mathrm{e}^{-i\mu t}$,
and rescale the dimensionful variables as follows:
\begin{eqnarray}
    r &  =& \tilde r \; \left ( \frac{1}{m} \right ), \quad
      \Phi  = \tilde \Phi \; (4\pi \; G_N)^{-1/2} , \cr
             \mu & =& \tilde \mu \; m, \quad \quad \quad 
                   \lambda  = \tilde \lambda \; 
                             (4\pi \; G_Nf^2),
\end{eqnarray}
The Einstein-Klein-Gordon system can then be written with
dimensionless variables. 
\begin{align}
  \label{eq:KG-eom-rescaled}
  &
    \left ( \frac{\tilde \mu^2}{B} + 1 \right )  \tilde
    \Phi^2 
    +  \frac{1}{A}  {\tilde \Phi^{\prime 2}}
    + \frac{1}{2}\tilde \lambda \tilde
    \Phi^4    - \frac{A'}{\tilde r  A^2} +
    \frac{1}{\tilde r^2 A} -
    \frac{1}{\tilde r^2} = 0, 
  \cr
& \left ( \frac{\tilde \mu ^2 }{B}  - 1 \right )\tilde \Phi^2
  +  \frac{1}{A} \tilde  \Phi'^2
        - \frac{1}{2} \tilde \lambda \tilde \Phi^4
-\frac{B'}{\tilde r A B} -
        \frac{1}{\tilde r^2 A} + \frac{1}{\tilde r^2}
        = 0,
        \cr
  &
    \frac{1}{A}{\tilde  \Phi^{\prime\prime }}
    +\left (\frac{\tilde \mu^2}{B} - 1 \right )\tilde \Phi +
  \tilde \Phi' \left (\frac{B'}{2 AB} - \frac{A'}{2A^2} +
  \frac{2}{A \tilde r}\right )  - \tilde \lambda \tilde \Phi^3 = 0,
\end{align}
Note that all derivatives are taken with respect to the rescaled
variables. From the above, we can solve the wave function and the
background geometry simultaneously. 

\subsection{The Boson Star Mass Profile}
\label{sec:boson-star-mass}
After the wave function and the metric are solved, the ADM mass of the boson star is
computed as
\begin{align}
  M_{\text{BS}} &=
      \int_0^\infty d^3x \;  T^0_0
      \cr
  & =
    \int_0^\infty dr  \; 4\pi r^2\;
    \left (
    \frac{\mu^2}{2B}\Phi^2 
    +\frac{1}{2}m^2 \Phi^2
    +\frac{1}{2A}\partial_r \Phi^2
    +\frac{1}{4}\lambda \Phi^4   \right ).
    \cr
    &
\end{align}

The extended nature of the boson wave function implies that its surface has to be defined as a sphere which contains a given percentage of its total mass. We choose the radius $R_{90}$ that encloses $90\%$ of the total mass. The star mass is denoted as $M_{\text{BS}}$. 

\begin{figure}[t]
  \centering
  \includegraphics[width=.8 \textwidth]{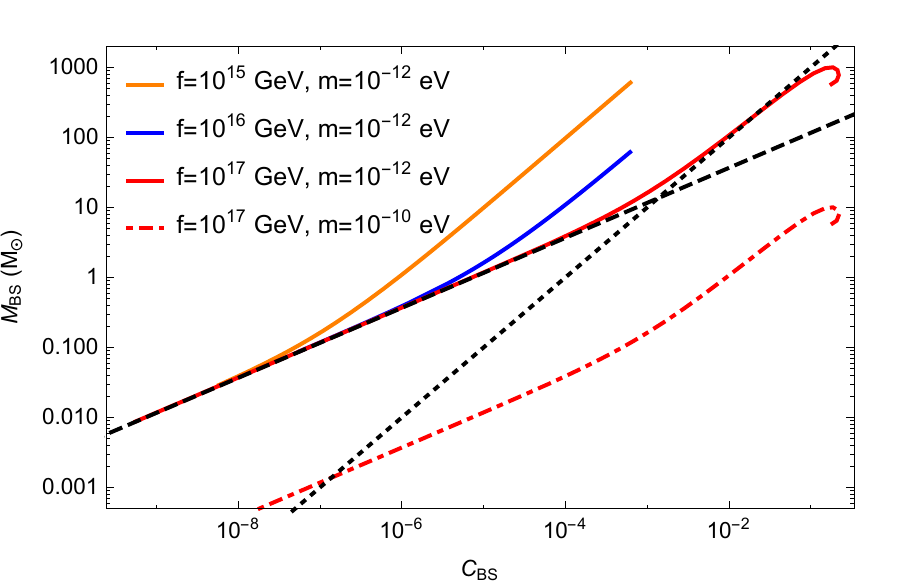
  }
  \caption{The profile of boson stars resulting from the scalar theory
    shown in Eq.~\eqref{eq:Lagrangian-KG-curved-space}. The mass
    curves are results from solving Eq.~\eqref{eq:KG-eom-rescaled}
    numerically.  Each curve corresponds to a single theory point
    $(m, f)$. All three solid curves have the same scalar mass,
    $m=10^{-12}\;\mathrm{eV}$, and differ in $f$. The red dot-dashed
    curve corresponds to
    $f = 10^{17}\; \mathrm{GeV},
    m=10^{-10}\;\mathrm{eV}$
    .  Note that we solve the curves of $f = 10^{17}\; \mathrm{GeV}$ up
    to the unstable regime (the spiral end where the star mass
    decreases,) and only show the relevant stable branch for
    $f = 10^{16}\;\mathrm{GeV}$ and $f = 10^{15}\;\mathrm{GeV}$. The dashed curve corresponds to the slope of
    $M_{\text{BS}}\propto C_{\text{BS}}^{1/2}$. The dotted curve
    corresponds to $M_{\text{BS}}\propto C_{\text{BS}}$. See the main
    text for more details. 
} 
  \label{fig:bs-mass-profile}
\end{figure}

The boson stars parameters $M_{\text{BS}}$ and $R_{90}$ are not independent
  for a given particle theory. For a given central density $\phi(0)=x$,
the wave function can be solved uniquely, leading to a pair of
solutions $(M_{\text{BS}}(x), R_{90}(x))$. Varying the central density $\phi(0)$
yields a curve in the $M_{\text{BS}}$-$R_{90}$ plane, which is the boson star mass profile.
The gravitational wave signals that we will be interested in are
largely affected by the compactness of the inspiralling object.  We
define the compactness as
$C_{\text{BS}}=G_N M_{\text{BS}}/(R_{90} c^2)$ in SI units,
and show the star mass profile in the $C_{\text{BS}}$-$M_{\text{BS}}$ plane in Fig.~\ref{fig:bs-mass-profile}. 

\subsection{Linear vs Nonlinear Regime}
\label{sec:linear-vs-nonlinear}

We notice from Fig.~\ref{fig:bs-mass-profile} that the stable branch of the boson star admits two regimes where 
$M_{\text{BS}}$ and $C_{\text{BS}}$ scale differently. In the linear regime, the boson star mass scales as 
$M_{\text{BS}}\propto C_{\text{BS}}^{1/2}$, while in the nonlinear regime, the scaling goes as 
$M_{\text{BS}}\propto C_{\text{BS}}$. These scalings can be understood as follows.
In the non-relativistic limit, the following two parameter ansatz can be taken
\begin{align}
  \Phi \approx \sqrt{\frac{N}{\pi mR^3}}\mathrm{e}^{-r/R}.    
\end{align}
The boson star energy is approximated as
\begin{align}
  H(N,R) & \approx
      \frac{N}{2m R^2}       + \frac{\lambda N^2}{32 \pi f^2 R^3}          - \frac{5 G_Nm^2N^2}{16 R}.
     \label{eq:HNR}  
\end{align}
In the regime of linear scaling, the gradient (first term) balances with gravity (last term). This leads to $N\sim 1/(m^3R)$, which is
\begin{align}
  C_{\text{BS}} \sim \frac{m N}{R} \sim m^2 M_{\text{BS}}^2. 
\end{align}
For a fixed compactness, the mass of the boson star is inversely
proportional to the scalar mass, $M_{\text{BS}}\propto 1/m$; for a
fixed scalar mass, the compactness is proportional to
  the  boson star mass squared, $M_{\text{BS}}\propto C_{\text{BS}}^{1/2}$.

In the regime of nonlinear scaling, it is the repulsive term (second term) that balances with gravity (last term) in Eq.~\eqref{eq:HNR}. This leads to $R\sim 1/(mf)$, which indicates
\begin{align}
  C_{\text{BS}} \sim \frac{M_{\text{BS}}}{R} \sim m\cdot f\cdot M_{\text{BS}}. 
\end{align}
For a given compactness, the star mass is affected by both $m$ and
$f$, with the relation $M_{\text{BS}}\propto 1/(mf)$; for a given
theory point $(m, f)$, the mass of the boson star scales with
compactness linearly, $M_{\text{BS}}\propto C_{\text{BS}}$.

The above behavior in both linear and nonlinear regimes are observed in the numerical solutions, shown in Fig.~\ref{fig:bs-mass-profile}.

\section{\label{sec:EMRI}Extreme Mass Ratio Inspiral and Gravitational Waves}

Having obtained the profile of the boson star, we now turn to a calculation of the gravitational waves from EMRI.

\subsection{Stellar Dynamics near the SMBH and EMRI}

A SMBH resides in the center of most galaxies and its mass ranges from $10^6 M_{\odot}$ to 
$10^9 M_{\odot}$. Compared with the mass of a typical galaxy hosting
it ($10^{11} M_{\odot}$), the mass
of the SMBH is negligibly small.
It is only the innermost region of the galaxy, within several parsecs around the SMBH, where the SMBH
can have a dominant effect. It is here that an EMRI system can be formed.
This region is defined by the radius of influence of the SMBH:
\begin{equation}\label{eq:RoI}
  r_h = \frac{G M}{\sigma^2}  = 2 \text{pc} \left( \frac{M}{3 \times 10^6 M_{\odot}}
  \right)^{1/2} ,
\end{equation}
where $\sigma$ is the velocity dispersion in the galactic 
bulge and in the second equality above,
the $M - \sigma$ region is used~\cite{Ferrarese:2000se,Gebhardt:2000fk,Tremaine:2002js}. 

Within the radius of influence, the potential of the SMBH dominates 
and the total mass of the stellar objects is roughly equal to the mass
of the SMBH. Here, the stellar population consists of neutron stars,
stellar black holes, white dwarfs(see e.g. Ref.~\cite{Alexander:2005jz}) and possibly ECOs. This region is
very crowded, hence the ECO being within the SMBH
  influence radius does not guarantee a slow inspiral to be successfully finished. The main physical mechanisms that affect the inspiral of the ECO include
two-body relaxation and resonant relaxation. The studies of the dynamics of the stars rely on a 
phase space analysis (we refer the reader to Ref.~\cite{alexander:review} for a recent review on this subject). 
As the SMBH devours any compact objects that get sufficiently close to it, there is a loss cone in the phase
space. The slow inspiral of EMRI requires that the ECO be sufficiently close to the SMBH so that  
the time scale for the EMRI to complete is smaller than the typical relaxation time scales and the EMRI
can finish without being interrupted by the other stars. It is for this kind of ECO in the slow inspiral phase
that an EMRI be formed with gravitational waves detectable by LISA.

\subsection{Gravitational Waves}

The standard EMRI system consists of a stellar black hole of typical mass $10 M_{\odot}$ and a SMBH
of $10^6 M_{\odot}$, resulting in a mass ratio that is an extremely small (or large) number. Calculating the waveforms of the EMRI system is a technically challenging and ongoing effort. 
In Ref.~\cite{Babak:2017tow}, three different waveforms have been used to tackle the data analysis issues for the EMRI systems. This  includes the Kludge-family waveforms, which started with the early 
work of Ref.~\cite{Barack:2003fp} and is now called the Analytical Kludge (AK) model. This has subsequently grown into the numerical AK~\cite{Babak:2006uv} and augmented AK~\cite{Chua:2017ujo} models. Also considered in Ref.~\cite{Babak:2017tow} is the result presented in Ref.~\cite{Finn:2000sy} for circular orbits in the 
equatorial plane, which is calculated based on the Teukolsky
formalism~\cite{Teukolsky:1973ha,Sasaki:1981sx}. It was found that the results of Ref.~\cite{Finn:2000sy}
show overall consistency with the other waveforms.~\footnote{The
  emission of gravitational waves tends to render the orbit of the
  compact object circular. However astrophysical studies still suggest
  that the compact object can have significant eccentricity such as
  through induction \cite{Randall:2017jop}.} We thus assume, for simplicity, that the compact object has a circular orbit in the equatorial plane. 

The gravitational wave amplitude is described in terms of the dimensionless characteristic strain, which is decomposed
into a set of harmonics labelled by $m$:
\begin{eqnarray}\label{eq:CharStrain}
&&h_{c,1} = \frac{5}{\sqrt{672\pi}} \frac{\eta^{1/2}M}{r_o} \tilde{\Omega}^{1/6} \mathcal{H}_{c,1}\ ,
  \nonumber \\
&&h_{c,m} = \sqrt{\frac{5(m+1)(m+2)(2m+1)! m^{2m}}{12\pi (m-1) [2^m
m!(2m+1)!!]^2}}
\frac{\eta^{1/2}M}{r_o}  \times \tilde{\Omega}^{(2m-5)/6}
\mathcal{H}_{c,m}, \quad m\ge 2\ . \label{eq:hcm}
\end{eqnarray}
where $\eta = M_{\text{BS}}/M$ is the mass ratio of boson star ($M_{\text{BS}}$) and SMBH ($M$); 
$r_o$ is the distance to the Earth;  $\tilde{\Omega} \equiv M \Omega = 1/(\tilde{r}^{3/2}+a)$
is a dimensionless orbital angular frequency where $\tilde{r} \equiv r/M$ with $r$ the 
Boyer-Lindquist radial coordinates of the orbit; 
$\mathcal{H}_{c,m}$ captures the relativistic corrections.
~\footnote{We use geometrized units here, i.e. $G_N=1$ and $c=1$.}

Given the above gravitational strain, the detectability of the EMRI gravitational waves signal is quantified by the signal-to-noise ratio (SNR):
\begin{eqnarray}\label{eq:SNR}
\text{SNR}^2 = \sum_{m} \int
\left[ \frac{h_{c,m}(f_m)}{h_n(f_m)} \right]^2 d \ln f_m,
\label{eq:snr}
\end{eqnarray}
which is obtained with matched-filtering~\cite{Moore:2014lga}. We
choose $\text{SNR} = 20$ as the threshold for detection~\cite{Babak:2017tow}, though a lower value of $15$
has also been demonstrated to work~\cite{Babak:2009cj}.

\begin{figure}[t]
\centering
\includegraphics[width=0.7\textwidth]{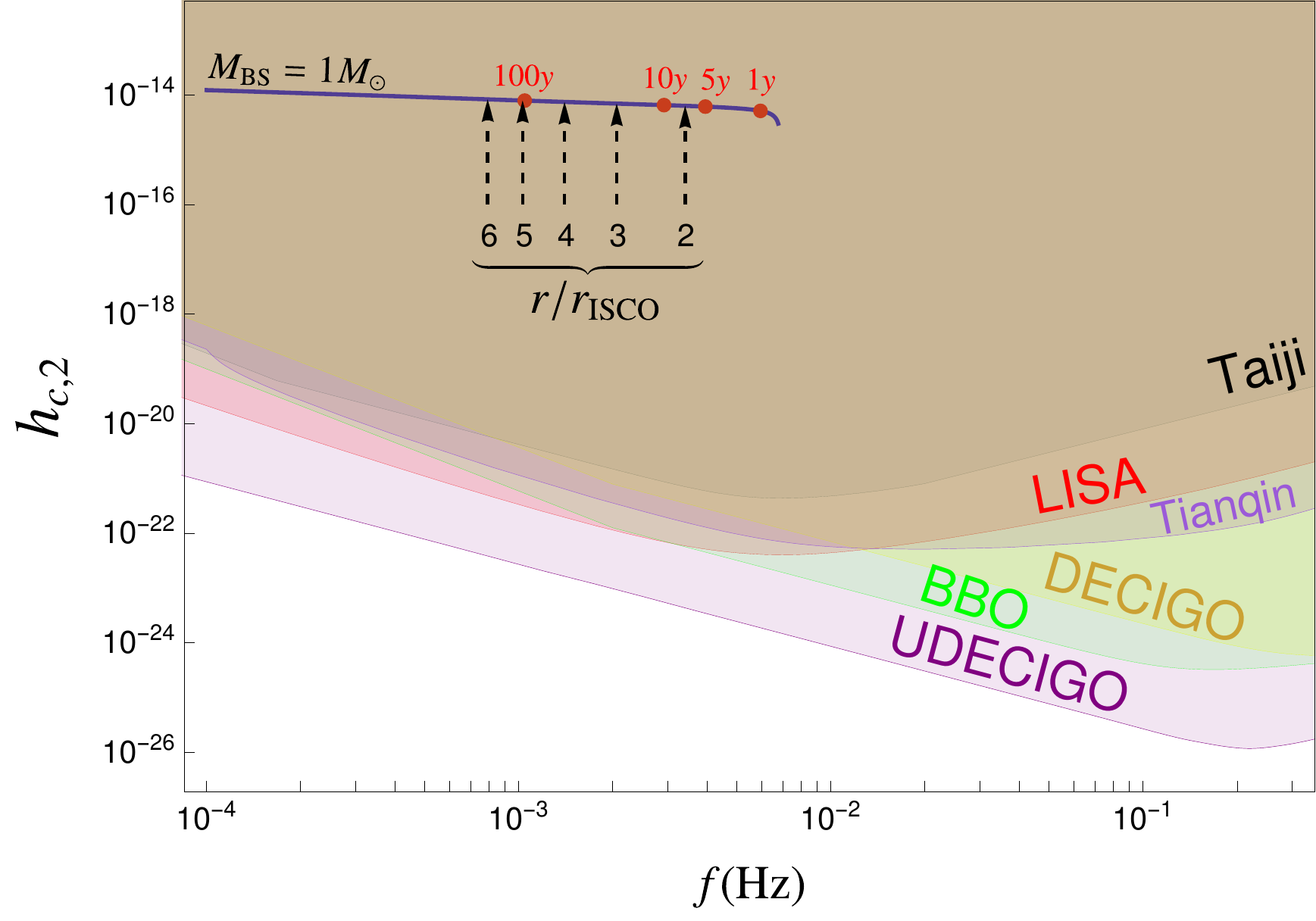
}
\caption{
\label{fig:hc}
The gravitational wave characteristic strain (blue solid line) from an EMRI system for a boson star with $M_{\text{BS}} = 1 M_{\odot}$ orbiting a super massive black hole with $M_{\text{SMBH}} = 4.1\times 10^6 M_{\odot}$, $a=0.999$ and a distance
  to Earth same to that of the Sgr A$^{\ast}$. The dots on this line denote the time remaining to the ISCO for several choices of time stamps as labeled in the text. The vertical dashed lines show the locations where the radial distance to ISCO is at several multiples of $r_{\text{ISCO}}$. The lower border of the color-shaded regions are 
the sensitivity curves of several proposed future space-based interferometer gravitational wave detectors.
}
\end{figure}

In the above equation, the range of the frequency to be integrated over should be taken as the overlap between the band of the detector and of the signal during the time when the detector is operating. For simplicity, we choose the upper end of the frequency band as the one at the ISCO when the compact object remains intact as it enters the event horizon of the SMBH. Conversely, when tidal disruption happens outside the horizon (to be discussed in the following section) the upper end of the band is chosen to be the frequency at the tidal radius of the compact object. The radius at ISCO for a SMBH with mass $M$ and spin $a$ has been given in Ref.~\cite{Bardeen:1972fi}. 

Another important quantity in our calculations is the time remaining to ISCO. For a given value of the gravitational wave frequency, it is given by~\cite{Finn:2000sy}
\begin{eqnarray}
T = \frac{5}{256} \frac{1}{\eta} \frac{M}{\tilde{\Omega}^{8/3}} \mathcal{T},
\end{eqnarray}
where $\mathcal{T}$ is a general relativistic correction factor, which is a function of the frequency and the SMBH spin. This equation can be reversed to calculate the frequency, given the time remaining to 
ISCO. For example, setting $T$ to be $5$ years gives the lower end of the frequency band in the SNR for the fiducial detector of LISA in the case that the compact object enters the horizon intact.

Since the boson star lingers around the ISCO for a very long time, it can travel a large number of cycles around the SMBH. This number is given by~\cite{Finn:2000sy}
\begin{eqnarray} 
N = \frac{1}{2\pi} \int d \Phi = \frac{1}{64 \pi} \frac{1}{\eta \tilde{\Omega}^{5/3}} \mathcal{N},
\end{eqnarray} 
where $\mathcal{N}$ captures the general relativistic correction, similar to the previous quantity 
$\mathcal{T}$. $\mathcal{N}$ equals one as $r \rightarrow \infty$ and vanishes as 
$r \rightarrow r_{\text{ISCO}}$. A typical value for $N$ in LISA band is $10^6$, much larger than the 
corresponding quantity in the LIGO case.
\begin{figure}[t]
\centering
\includegraphics[width=0.466\textwidth]{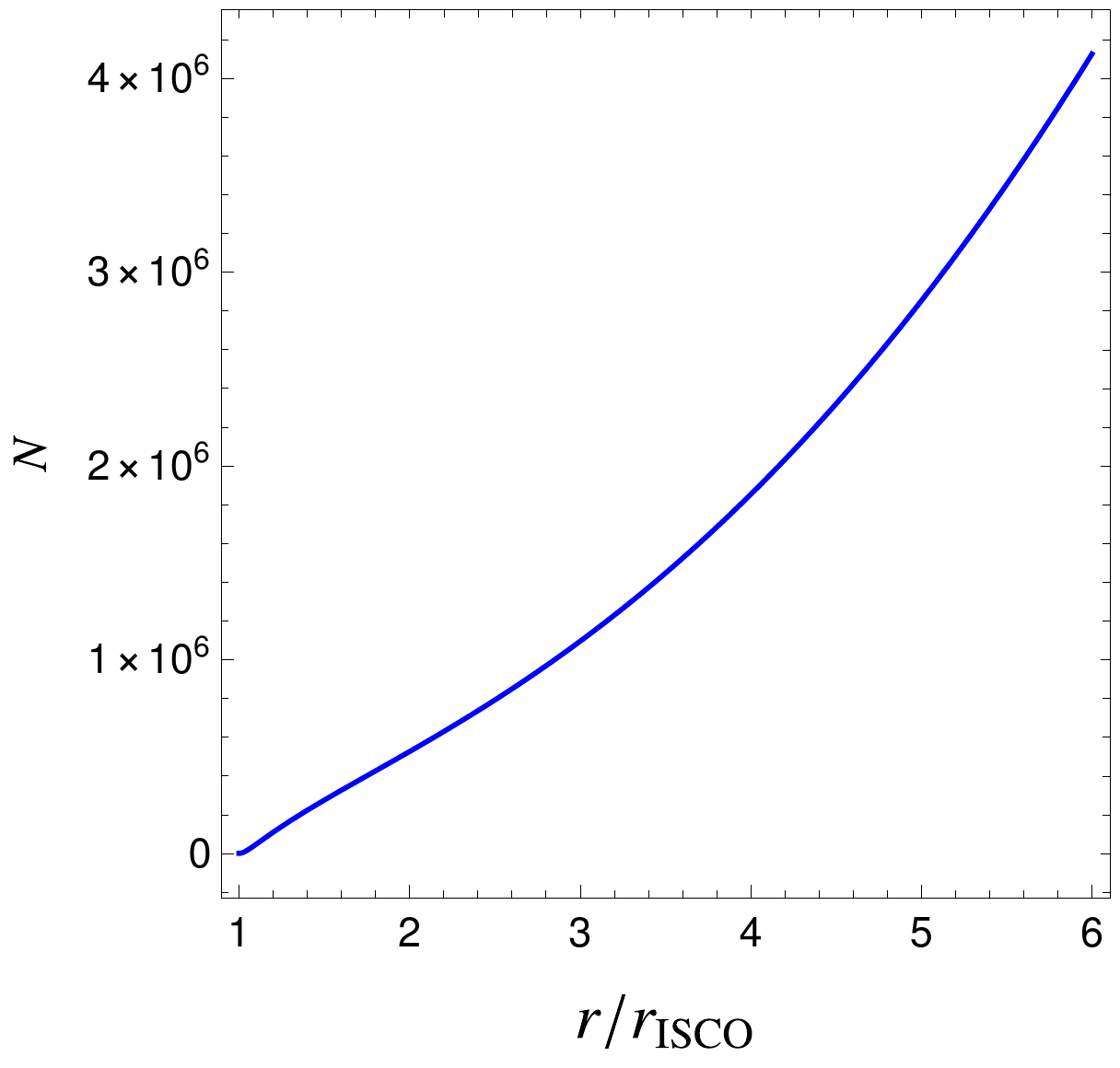
}
\quad
\includegraphics[width=0.45\textwidth]{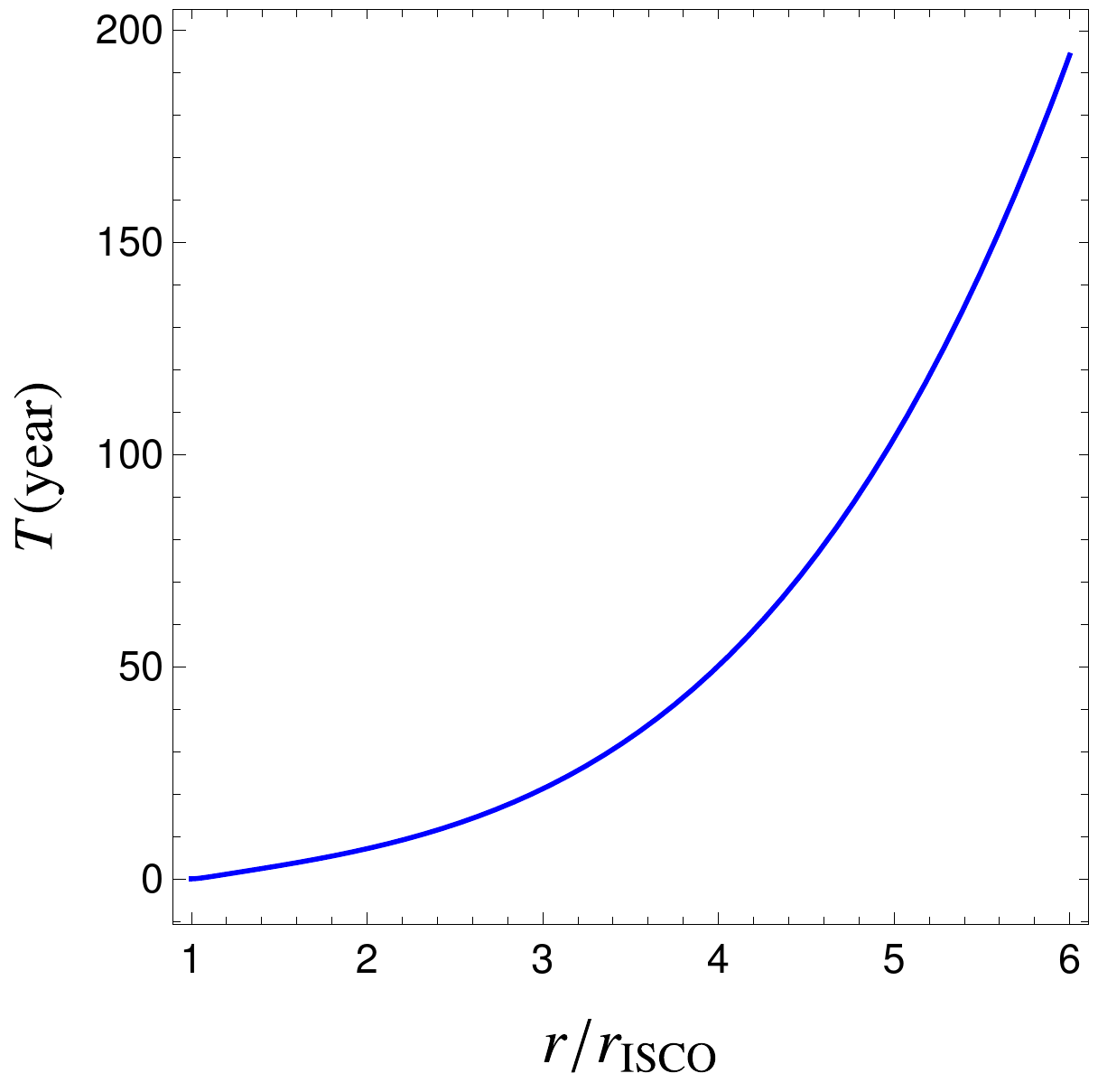
}
\caption{
\label{fig:N}
The number of circles remaining (left) and the time remaining (right) till the ISCO
as a function of the distance from the SMBH $r$ normalized by the ISCO radius,
for the same EMRI system as in Fig.~\ref{fig:hc}.
}
\end{figure}
Since the SNR is proportional to the square root of $N$~\cite{Moore:2014lga}, 
it is greatly enhanced for the EMRI system. To phrase this in an illuminating way, 
LISA will be able to see much further distances. 
\begin{figure}[t]
\centering
\includegraphics[width=0.85\textwidth]{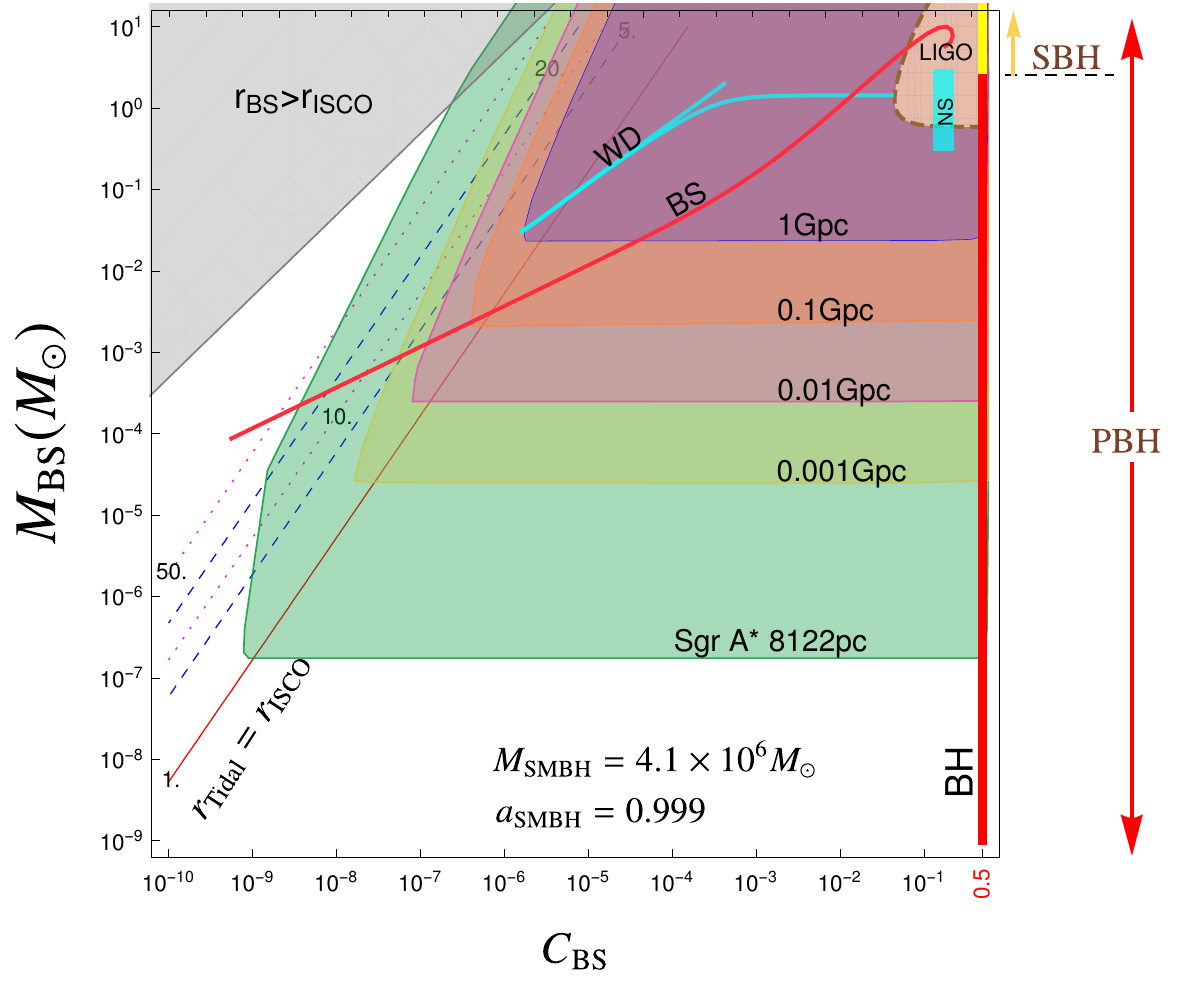
}
\caption{
\label{fig:mc}
The regions in the compactness-mass plane of boson stars that can be probed by LISA and LIGO. The upper-right brown 
corner shows the 
region where the gravitational waves from a equal mass boson star binary can be detected by LIGO for a distance
to Earth larger than $0.2$ Gpc, assuming the detection threshold of
the SNR to be 8. 
The other color-shaded regions denote the parameter space for which an EMRI system, composed of a 
boson star (with corresponding compactness and mass) and a supermassive black hole (with the mass and spin 
  as in Fig.~\ref{fig:hc}), can produce gravitational waves with a threshold SNR of 20 for the LISA detector assuming 
a mission time of 5 years. The different colors denote the different maximum distances where such detection can be made 
and with the maximum distances corresponding to the outer boundaries labeled by the associated text.
The straight lines show the tidal radius  in units of $r_{\text{ISCO}}$, and the one where $r_{\text{Tidal}}$ coincides with the ISCO is shown by the red line. The gray region in the top left corner gives a boson star radius 
that is larger than the radius of ISCO. The red curve is a representative boson star solution for the parameter choice of $m_{\phi}=10^{-10}\text{eV}$ and $f=10^{17}\text{GeV}$. Also plotted are the regions of several species of other compact
  objects, such as white dwarfs (WD), neutron stars (NS) and black holes (BH) (stellar black holes (SBH) by  vertical yellow line or primordial black holes (PBH) by vertical red line.)
}
\end{figure}
Moreover, due to the large circle time recorded over several years, the analysis of the gravitational wave waveforms from an EMRI system can lead to a very precise determination of the parameters in the system, such as the masses of the system and the SMBH 
spin. The uncertainties can be as small as $10^{-5}$~\cite{Barack:2003fp}. Therefore detection of gravitational waves from an EMRI system where the smaller object has sub-solar mass would definitely rule out the possibility of it being stellar black holes, neutron stars and white dwarfs.

As an example, the solid blue line  in Fig.~\ref{fig:hc} shows the gravitational wave characteristic strain as a function of the frequency for a $1 M_{\odot}$ boson star inspiralling into the SMBH of mass $4.1 \times 10^6 M_{\odot}$(see caption for
 more details). The dots on this line denote several time stamps before reaching the ISCO. Also plotted in Fig.~\ref{fig:hc} are experimental sensitive regions for several proposed detectors, including LISA with configuration C1~\cite{Klein:2015hvg}, the Taiji~\cite{Gong:2014mca} and Tianqin 
projects~\cite{Luo:2015ght}, the Big Bang Observer (BBO), the DECi-hertz Interferometer
Gravitational wave Observatory (DECIGO)~\cite{Moore:2014lga}, and 
Ultimate-DECIGO (UDECIGO)~\cite{Kudoh:2005as}. In addition, the number of circles remaining and the 
time remaining until the ISCO is shown in the left and right panels of Fig.~\ref{fig:N} respectively, 
as a function of $r/r_{\text{ISCO}}$. We see that the typical size of the number of circles remaining in the LISA band is indeed very large.

\section{Results}
\label{sec:results}

In this section, we present our results. We first show the results of our EMRI analysis on the mass versus compactness plane of boson stars. We then discuss the constraints in theory space. We finally comment on the event rate and the possibility of correlated electromagnetic signals.

\subsection{Constraints on Mass versus Compactness}

The EMRI analysis described in the previous section leads to strong constraints on the properties of boson stars, which we now describe.
Our results are presented in Fig.~\ref{fig:mc}, where we show the regions in the  mass versus compactness plane of boson 
stars that can be probed by LISA and LIGO (see caption for
details). 
Gravitational waves from a boson star
binary that can be detected by LIGO are obtained in the upper-right
corner, for large mass and compactness. For the LIGO bounds, a
distance to Earth larger than $0.2$ Gpc is taken, and the detection
threshold of the SNR is assumed to be 8 \cite{Dominik:2014yma}.

The color-shaded regions in Fig.~\ref{fig:mc} correspond to an EMRI system emitting gravitational waves with SNR larger than 20 for the LISA detector assuming 5 years of mission time. The EMRI system is composed of a boson star with the corresponding compactness and mass, and a SMBH of mass $M_{\text{SMBH}}=4.1 \times 10^6 M_{\odot}$ and spin $a = 0.999$, the same
as that used in Fig.~\ref{fig:hc}. The different colors denote different distance ranges where such detection can be made. The red curve is a representative boson star solution for the parameter choice of $m_{\phi}=10^{-10}\text{eV}$ and $f=10^{17}\text{GeV}$. Also plotted are the regions of white dwarfs (WD), neutron stars (NS) and black holes (BH) (stellar black holes (SBH) by  vertical yellow line or primordial black holes (PBH) by vertical red line).

\subsection{Tidal Radius}
\label{sec:tidal-radius} 
We now turn to a discussion of the tidal radius.
For an EMRI system where the small object is a white dwarf, neutron star or primordial black hole, the object is compact enough so 
that it can generally pass the ISCO without being tidally disrupted, resulting in a gravitational wave signal that lasts until 
the ISCO. 
Boson stars, on the other hand, can have compactness
 well below $0.5$ yet still are detectable at LISA. For a boson star with a fixed mass, decreasing the compactness leads to an increase in the tidal radius. As the tidal radius reaches values greater than the ISCO, the boson star will be tidally disrupted outside the ISCO. The gravitational wave signal will then be cut off at the tidal  radius and the maximal frequency recorded by the detector will be smaller than what can be achieved at the ISCO. 

We therefore see that for boson stars, the frequency range in the SNR needs to take the tidal disruption into account. Thus for each choice of $(M_{\text{BS}}, \cbs)$, we need to find the corresponding tidal radius and
compare it with $r_{\text{ISCO}}$ to determine the correct frequency band to be used in the SNR calculations.

For a non-spinning SMBH with mass $M$, the tidal radius can be obtained simply by equating the force of the SMBH at $r_{\text{Tidal}}$ with the self-gravitating force of the BS. This gives
the well known result for $\rtidal$~\footnote{We note that for a Kerr SMBH, a more precise definition of the
tidal radius can be found by calculating the general relativistic 
tidal tensor (see Ref.~\cite{stone:2015tidal} for a review). 
We leave a more detailed investigation of EMRI with spinning SMBHs for a 
future publication.}:
\begin{eqnarray}
  r_{\text{Tidal}} = R_{90} \left(\frac{M}{M_{\text{BS}}}\right)^{1/3} = \frac{(\mbs^2 M)}{\cbs}^{1/3} ,
\end{eqnarray}
where in the second step we have traded $\rbs$ for $\cbs$. Note that this radius is roughly the same as the Roche radius. Since $r_{\text{Tidal}} \propto M^{1/3}$ and $r_{\text{ISCO}} \propto M$, the tidal radius will be inside the horizon for SMBH with mass larger than a threshold, known as the Hill mass~\cite{stone:2015tidal}.

In Fig.~\ref{fig:mc}, several choices of the tidal radius are shown in units of $r_{\text{ISCO}}$. Since this is a log-log plot, each $r_{\text{Tical}}$ appears as a straight line. The straight lines show the tidal radius in units of $r_{\text{ISCO}}$, and the one where $r_{\text{Tidal}}$ coincides with the  ISCO is shown by the red line. The gray region in the top left corner gives a boson star radius that is larger than the radius of ISCO. 

With $\rtidal$ determined above, the next step is to find the
corresponding frequency at $\rtidal$ and the associated frequency 5
years before reaching $\rtidal$, which is used as the lower limit of
the frequency band. Therefore, for the boson stars that experience
tidal disruption, the frequency band of the gravitational wave signal
is $(f(T_{\text{Tidal}}+5\text{year}), f(T_\text{Tidal}))$ instead of
$(f(T_{\text{ISCO}}+5\text{year}), f(T_\text{ISCO}))$ because
$T_\mathrm{Tidal}> T_\mathrm{ISCO}$.

For a Kerr SMBH, the ISCO radius decreases as its spin parameter increases from negative (retrograde motion)
to positive values (prograde motion). Therefore, an ECO can in general get closer towards the SMBH for prograde orbits 
before it reaches the larger of $r_\mathrm{ISCO}$ and
  $r_\mathrm{Tidal}$, 
compared with the retrograde case whose $r_{\text{ISCO}}$ is larger.
Conversely, as the SMBH spin decreases, $r_{\text{ISCO}}$ expands and
the condition $r_{\text{ISCO}} = r_{\text{Tidal}}$ will be reached for
less compact BS when its mass is fixed, which means the boson
  stars are less likely to get tidally disrupted. This corresponds to
that the
straight lines in Fig.~\ref{fig:mc} will shift towards the left.

We need to ensure that the tidal radius is smaller than the radius of influence of the SMBH, so that an EMRI is formed in the first place. This condition is generally straightforward to satisfy for most cases. For example, for a SMBH with mass $4.1 \times 10^6 M_{\odot}$, the tidal radius needs to be as large as $(10^6 \sim 10^7) r_{\text{ISCO}}$ to become comparable to the radius of influence $\sim 1$ pc. Therefore boson stars that generate gravitational waves within the LISA band are certain to be within the radius of influence. This can be seen in Fig.~\ref{fig:mc}, where the regions of parameter space probed by LISA can only be reached for tidal radius of $\mathcal{O}(10) r_{\text{ISCO}}$.

Besides the tidal radius which affects less compact boson stars, other tidal interactions can show up, modifying boson star profiles as well as the resulting gravitational wave waveforms. A boson star can be tidally deformed before it reaches the tidal radius. The deformability can be characterized by the tidal love numbers and has been studied for neutron stars~\cite{Damour:2009vw} and  boson stars~\cite{Cardoso:2017cfl,Binnington:2009bb}. The resulting impact on the gravitational wave waveforms, however, is a secondary effect~\cite{Barack:2003fp}. 

The same is true for the spin of the boson star, as well as the spin-orbital coupling effect. Astrophysical studies of the tidal interactions of the stars near the SMBH show that stars can accumulate spins through constantly squeezing at the pericenter. The same is expected to happen  for boson stars and a rotating boson star may be formed. Rotating boson star solutions have been considered in Ref.~\cite{Mielke:2016war}~\footnote{ The rotating boson star may have a different mass-radius relation compared to the non-rotating case. The main effect is in changing the tidal radius.}.  However as far as gravitational waves are concerned, these effects tend to be negligible.

We point out  that tidal effects in inspiralling binary systems have been studied as a potential discriminant between ordinary compact objects and boson stars \cite{Sennett:2017etc}.

\subsection{Probed Theory Space}
\label{sec:prob-theory-param}
\begin{figure}[t]
  \centering
  \includegraphics[width=.8\textwidth]{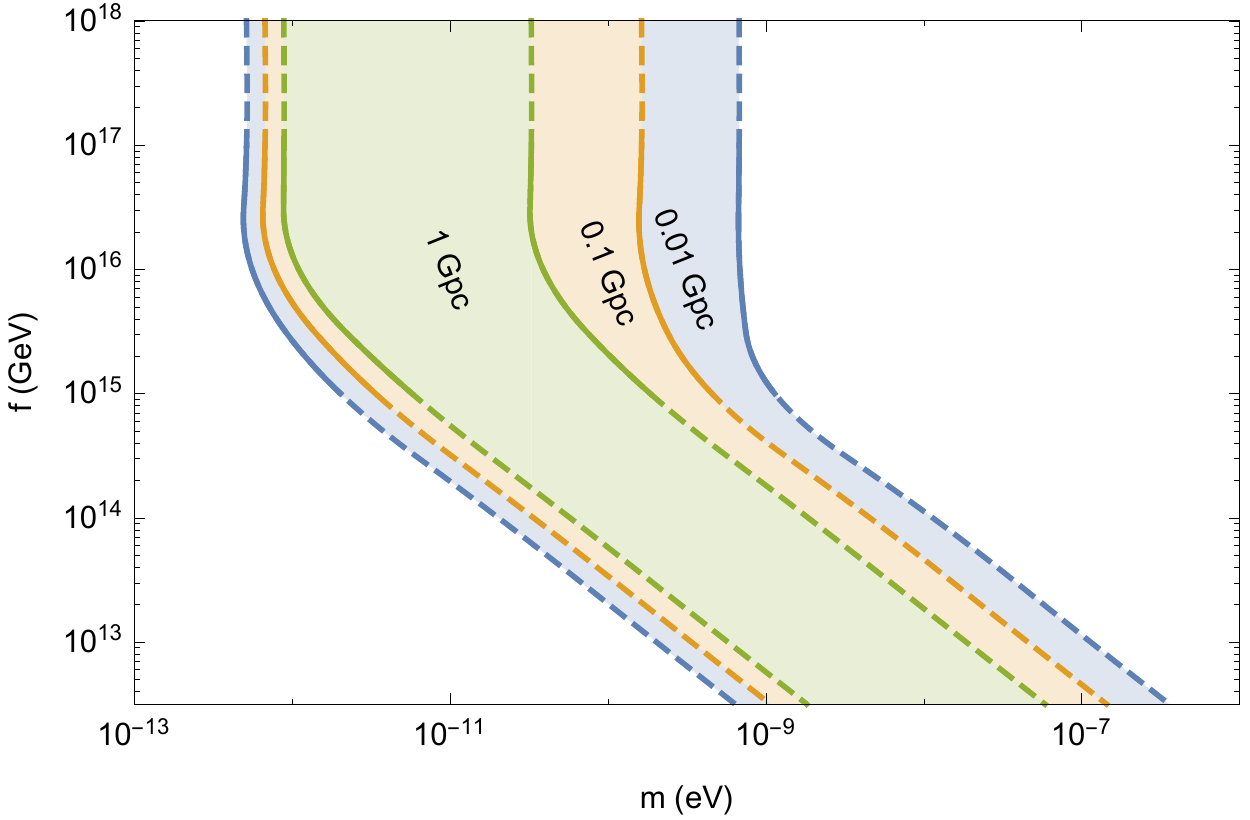
  }
  \caption{Parameter region which admits distinguishable boson
    stars. The right edges of the color shaded regions correspond to
    the mass profile being above the intersection point of
    $r_\mathrm{Tidal} = r_\mathrm{ISCO} $ and the LISA contours in
    Fig.~\ref{fig:mc}. The left edge corresponds to the requirement of
    a segment of the boson mass curve overlaps with LISA sensitivity
    band and has mass smaller than 1~$M_{\odot}$.
    The solid curves correspond to the transition region between the
    linear regime and the nonlinear regime, and are solved
    numerically. The dashed curve are computed analytically based on
    the numerical evidence discussed in
    Section~\ref{sec:linear-vs-nonlinear}. See the main text for
    definition of distinguishable boson stars.  }
  \label{fig:parameter-region}
\end{figure}

Based on the EMRI analysis, the probed region on the mass versus
compactness parameter space can now be translated into the theory
space $(m, f)$. Instead of translating all the probed
region to the theory parameter space, we look at a subset where boson
stars experience tidal disruption right before reaching $r_{ISCO}$.
Because the early termination of the inspiral signal gives us extra
information about the compactness of the infalling object, we refer
such boson stars as distinguishable boson stars.

The distinguishable boson star compactness corresponding to the
situation where tidal forces severely disrupt it right before ISCO is
given by the line of $r_\mathrm{Tidal} = r_\mathrm{ISCO}$ in
Fig.~\ref{fig:mc}. On the other hand, the infalling star needs to be
large enough to accumulate enough SNR at LISA. Therefore, for a given
luminosity distance, we require the boson star mass profile to be
above the intersection point of $r_\mathrm{Tidal} = r_\mathrm{ISCO}$
line and the sensitivity contour, which guarantees a segment of the
mass curve falling into the sensitivity region of LISA yet still has
an early termination feature. For the top three benchmark contours in
Fig.~\ref{fig:mc}, 
the intersection points correspond to
\begin{align}
  D_L & = 0.01\; \mathrm{Gpc}, & C_\mathrm{BS}  = 2.1 \times 10^{-7}, \;M_\mathrm{BS} = 2.5
        \times 10^{-4} \; M_\odot, \cr
        D_L & = 0.1\; \mathrm{Gpc}, & C_\mathrm{BS}  = 8.8 \times 10^{-7},\; M_\mathrm{BS} = 2.1
              \times 10^{-3} \; M_\odot, \cr
              D_L & = 1\; \mathrm{Gpc}, & C_\mathrm{BS} = 4.4 \times 10^{-6},\; M_\mathrm{BS} = 2.4
                    \times 10^{-2} \; M_\odot. 
\end{align}
Requiring the boson star mass profile being above these points
leads to the upper limit of the scalar mass in the theory space
shown in Fig.~\ref{fig:parameter-region}.  In addition, we require the
boson star mass curve has a segment such that a) it has overlap with LISA
sensitivity region; b) lighter than 1~$M_\odot$. This results in the
lower bound on the scalar mass that corresponds to the left edge of
Fig.~\ref{fig:parameter-region}.  We compute the region of transition
between linear regime and nonlinear regime numerically to get the
solid part of the curves, and invoke the argument in
Section~\ref{sec:linear-vs-nonlinear} to compute other regions
analytically. The color shaded region show the parameter space that
allow for such distinguishable stars.



\subsection{Correlated Electromagnetic Signals}

We sketch some general ideas about the possibility of obtaining correlated gravitational wave and electromagnetic signals coming from boson star-SMBH EMRIs. Tidal disruption flares coming from ordinary stars near SMBH are an important target in the astrophysics community. We refer to \cite{Komossa:2015qya} for a  review on current observations and \cite{Lodato:2015aoa} for a review of the theory. Therefore such electromagnetic products coming from the tidal disruption of exotic compact 
objects is an important and relatively unexplored topic. The signals would depend on the coupling of the boson to Standard Model fields, and one would have to take into account the resulting effects on the boson star profile. While such modeling is outside the scope of our present work, we note that the possibility of unique tidal flares from boson stars would constitute a key piece of observational evidence about their existence.

\subsection{\label{sec:event}Event Rate}

The estimation of the event rate of the EMRI from the ECO depends on
its population in the radius of influence of the SMBH, which in turn
depends on its formation mechanism and how the ECOs sank into the
galactic center. Without a prior knowledge of these, an upper
  limit of the event rate can be set by assuming that the ECOs formed from
  material that was part of the dark matter halo. While we do not
give a solid connection between ECO and dark matter, there have
already been 
many studies of this scenario in the literature~\cite{Croon:2018ybs}. If ECOs
indeed form from the dark matter halo, then the event rate for the
EMRI from the ECOs can be related to the halo profile.


Here we briefly sketch a few factors that may affect the EMRI event
rate.  Since EMRI is formed in the galactic center, the event rate of
the EMRI depends on the halo density in the innermost parsec region of
the galaxy.  While collisionless dark matter simulations lead to cuspy
NFW profile, dwarf galaxies usually turns out to be cored, due to
baryonic feedback~\cite{Navarro:1996bv,Pontzen:2011ty}, or new physics
such as dark matter self-interaction~\cite{Tulin:2017ara} or fuzzy
dark matter~\cite{Schive:2014hza,Hui:2016ltb}. The accuracy in
modeling the central density
profile~\cite{Navarro:1995iw,Navarro:1996gj} will affect the estimate
of the EMRI event rate. On the other hand, without a good prior on the
central density, the measured EMRI event rate could shed light on the
environment in the very center of the galaxy. In addition, the
presence of the SMBH may lead to a spike of the dark
halo~\cite{Gondolo:1999ef,Sadeghian:2013laa,Sandick:2016zeg}, which
might be further enhanced by the spinning of the
SMBH~\cite{Ferrer:2017xwm}.  Therefore variations can arise due to the
different choices of the halo profiles.

For a given dark matter halo profile, the halo provides an
initial ECO population near $r_h$. Within the radius of influence, the
ECOs will be redistributed due to collisions (gravitational
encounters) with other stars within the potential well of the
SMBH. For a relaxed ECO population, a density cusp can be formed, in
analogy to the Bahcall-Wolf stellar
cusp~\cite{gahcall:1976aa}. Furthermore, due to the effect of mass
segregation~\cite{Hopman:2006xn,AmaroSeoane:2010bq,Alexander:2008tq},
heavier ECOs will sink deeper toward the SMBH while lighter ECOs will
be pushed outward, leading to an additional enhancement (reduction)
of the profile for heavy (light) ECOs at small $r$, respectively. This introduces a
dependence of the event rate on ECO mass. If we assume an NFW profile and that 
the ECO makes up the whole dark halo, then LISA, assuming 5 years of observation time, will be able to see $\sim 3000$ EMRIs for ECOs heavier than $\sim 0.03 M_{\odot}$ (but
lighter than $\sim 10 M_{\odot}$ so that there is a sufficient number
of ECOs for a given halo profile). This estimate is based on a similar
analysis for primordial black hole dark matter~\cite{Guo:2017njn}.




\section{Summary\label{sec:summary}}

In this paper, we have proposed to use gravitational waves from EMRIs as a new method to search for boson stars.  We first introduced the particle physics model and computed the profile of the boson star in the mass-compactness plane. The results of numerically solving the Einstein-Klein-Gordon system were displayed in Fig.~\ref{fig:bs-mass-profile}, after which we discussed the linear and nonlinear regimes of the stable branch.

In Section \ref{sec:EMRI}, we provided the calculation of the gravitational waves from EMRI. The gravitational wave characteristic strain relevant for us is displayed in Fig.~\ref{fig:hc}. One of the most important features of the EMRI system is that a boson star can linger around the ISCO for a very long time, implying that a large number of cycles $N \sim 10^6$ around the SMBH are possible. Since the SNR is proportional to $\sim \sqrt{N}$, it is greatly enhanced for an EMRI. The number of cycles remaining  as a function of the distance from the SMBH, normalized by the ISCO radius, is shown in Fig.~\ref{fig:N}.

Our main results are presented in Fig.~\ref{fig:mc}, where we show the regions in the  mass versus compactness plane of boson stars that can be probed by EMRI at LISA. Clearly, our methods are able to probe boson stars down to very sub-solar mass and compactness. We are  careful to incorporate the effect of tidal disruption before the boson star reaches the ISCO; in such cases the signal is cut off at the tidal radius and the maximal frequency recorded by the detector is smaller than what can be achieved at the ISCO. 

Due to the large number of cycles recorded, EMRIs can lead to a very
precise mass determination of the boson star and distinguish it from
standard astrophysical compact objects. From Fig.~\ref{fig:mc}, where
standard astrophysical compact objects are also shown, it is clear
that a precise determination of the mass of the participating body
will distinguish between ECOs and usual compact objects. Tidal effects
in inspiralling binary systems further break the mass degeneracy of  
boson stars with other ECOs candidates.   Possible correlated electromagnetic signals, can also serve as potential discriminants.

On the particle physics side, our main results are shown in
Fig.~\ref{fig:parameter-region}. We looked into the theory parameter
region that admits boson stars in the LISA sensitivity contour yet
experience tidal force disruption before reaching
$r_\mathrm{ISCO}$. At the same time, we require the boson stars 
to be lighter than
1~$M_\odot$. The result shows a large parameter region where such
boson stars are allowed.
 
There are several future directions to be considered. This work paves
the way for future more detailed analysis including improvements on
the following sides: consideration of more generic eccentric orbits
and using more accurate waveforms, studying changes to the waveform
due to tidal deformability of the ECO to distinguish different species
of ECOs, dedicated analysis of the tidal disruption events,
application to BS with more complicated structures and to other
species of ECOs, etc. In addition, when the coupling to the
  Standard Model particles are taken into account, richer
  phenomenology is expected, such as electromagnetic signals that
  accompany the merger events. We leave this for future study.

\section{Acknowledgments}

We thank Djuna Croon for discussions and JiJi Fan for reading the
manuscript. K. Sinha and H. Guo are supported by the U. S. Department
of Energy grant DE-SC0009956. C. Sun is supported in part by the
International Postdoctoral Fellowship funded by China Postdoctoral
Science Foundation, and NASA grant 80NSSC18K1010.

\bibliographystyle{utphys}

\bibliography{mybib}

\providecommand{\href}[2]{#2}\begingroup\raggedright\begin{thebibliography}{10}

\bibitem{Colpi:1986ye}
M.~Colpi, S.~L. Shapiro, and I.~Wasserman, ``{Boson Stars: Gravitational
  Equilibria of Selfinteracting Scalar Fields},''
\href{http://dx.doi.org/10.1103/PhysRevLett.57.2485}{{\em Phys. Rev. Lett.}
  {\bfseries 57} (1986) 2485--2488}.

\bibitem{Tkachev:1986tr}
I.~I. Tkachev, ``{Coherent scalar field oscillations forming compact
  astrophysical objects},'' {\em Sov. Astron. Lett.} {\bfseries 12} (1986)
  305--308.
[Pisma Astron. Zh.12,726(1986)].

\bibitem{vanderBij:1987gi}
J.~J. van~der Bij and M.~Gleiser, ``{Stars of Bosons with Nonminimal Energy
  Momentum Tensor},''
\href{http://dx.doi.org/10.1016/0370-2693(87)90221-8}{{\em Phys. Lett.}
  {\bfseries B194} (1987) 482--486}.

\bibitem{JetzerP}
P.~{Jetzer}, ``{Boson stars},''
  \href{http://dx.doi.org/10.1016/0370-1573(92)90123-H}{{\em Phys. Rept.}
  {\bfseries 220} (Nov., 1992) 163--227}.

\bibitem{Schunck:2003kk}
F.~E. Schunck and E.~W. Mielke, ``{General relativistic boson stars},''
  \href{http://dx.doi.org/10.1088/0264-9381/20/20/201}{{\em Class. Quant.
  Grav.} {\bfseries 20} (2003) R301--R356},
\href{http://arxiv.org/abs/0801.0307}{{\ttfamily arXiv:0801.0307 [astro-ph]}}.

\bibitem{Liebling:2012fv}
S.~L. Liebling and C.~Palenzuela, ``{Dynamical Boson Stars},''
  \href{http://dx.doi.org/10.12942/lrr-2012-6, 10.1007/s41114-017-0007-y}{{\em
  Living Rev. Rel.} {\bfseries 15} (2012) 6},
  \href{http://arxiv.org/abs/1202.5809}{{\ttfamily arXiv:1202.5809 [gr-qc]}}.
[Living Rev. Rel.20,no.1,5(2017)].

\bibitem{Abbott:2016blz}
{\bfseries Virgo, LIGO Scientific} Collaboration, B.~P. Abbott {\em et~al.},
  ``{Observation of Gravitational Waves from a Binary Black Hole Merger},''
  \href{http://dx.doi.org/10.1103/PhysRevLett.116.061102}{{\em Phys. Rev.
  Lett.} {\bfseries 116} no.~6, (2016) 061102},
\href{http://arxiv.org/abs/1602.03837}{{\ttfamily arXiv:1602.03837 [gr-qc]}}.

\bibitem{1989PhRvL..63.1199G}
M.~{Gleiser}, ``{Gravitational radiation from primordial solitons and
  soliton-star binaries},''
  \href{http://dx.doi.org/10.1103/PhysRevLett.63.1199}{{\em Physical Review
  Letters} {\bfseries 63} (Sept., 1989) 1199--1202}.

\bibitem{Palenzuela:2006wp}
C.~Palenzuela, I.~Olabarrieta, L.~Lehner, and S.~L. Liebling, ``{Head-on
  collisions of boson stars},''
  \href{http://dx.doi.org/10.1103/PhysRevD.75.064005}{{\em Phys. Rev.}
  {\bfseries D75} (2007) 064005},
\href{http://arxiv.org/abs/gr-qc/0612067}{{\ttfamily arXiv:gr-qc/0612067
  [gr-qc]}}.

\bibitem{Palenzuela:2007dm}
C.~Palenzuela, L.~Lehner, and S.~L. Liebling, ``{Orbital Dynamics of Binary
  Boson Star Systems},''
  \href{http://dx.doi.org/10.1103/PhysRevD.77.044036}{{\em Phys. Rev.}
  {\bfseries D77} (2008) 044036},
\href{http://arxiv.org/abs/0706.2435}{{\ttfamily arXiv:0706.2435 [gr-qc]}}.

\bibitem{Amin:2013eqa}
M.~A. Amin, E.~A. Lim, and I.-S. Yang, ``{A scattering theory of
  ultrarelativistic solitons},''
  \href{http://dx.doi.org/10.1103/PhysRevD.88.105024}{{\em Phys. Rev.}
  {\bfseries D88} no.~10, (2013) 105024},
\href{http://arxiv.org/abs/1308.0606}{{\ttfamily arXiv:1308.0606 [hep-th]}}.

\bibitem{Bezares:2018qwa}
M.~Bezares and C.~Palenzuela, ``{Gravitational Waves from Dark Boson Star
  binary mergers},'' \href{http://dx.doi.org/10.1088/1361-6382/aae87c}{{\em
  Class. Quant. Grav.} {\bfseries 35} no.~23, (2018) 234002},
\href{http://arxiv.org/abs/1808.10732}{{\ttfamily arXiv:1808.10732 [gr-qc]}}.

\bibitem{Helfer:2018vtq}
T.~Helfer, E.~A. Lim, M.~A.~G. Garcia, and M.~A. Amin, ``{Gravitational Wave
  Emission from Collisions of Compact Scalar Solitons},''
  \href{http://dx.doi.org/10.1103/PhysRevD.99.044046}{{\em Phys. Rev.}
  {\bfseries D99} no.~4, (2019) 044046},
\href{http://arxiv.org/abs/1802.06733}{{\ttfamily arXiv:1802.06733 [gr-qc]}}.

\bibitem{Sanchis-Gual:2018oui}
N.~Sanchis-Gual, C.~Herdeiro, J.~A. Font, E.~Radu, and F.~Di~Giovanni,
  ``{Head-on collisions and orbital mergers of Proca stars},''
  \href{http://dx.doi.org/10.1103/PhysRevD.99.024017}{{\em Phys. Rev.}
  {\bfseries D99} no.~2, (2019) 024017},
\href{http://arxiv.org/abs/1806.07779}{{\ttfamily arXiv:1806.07779 [gr-qc]}}.

\bibitem{Croon:2018ybs}
D.~Croon, J.~Fan, and C.~Sun, ``{Boson Star from Repulsive Light Scalars and
  Gravitational Waves},''
\href{http://arxiv.org/abs/1810.01420}{{\ttfamily arXiv:1810.01420 [hep-ph]}}.

\bibitem{Croon:2018ftb}
D.~Croon, M.~Gleiser, S.~Mohapatra, and C.~Sun, ``{Gravitational Radiation
  Background from Boson Star Binaries},''
  \href{http://dx.doi.org/10.1016/j.physletb.2018.03.055}{{\em Phys. Lett.}
  {\bfseries B783} (2018) 158--162},
\href{http://arxiv.org/abs/1802.08259}{{\ttfamily arXiv:1802.08259 [hep-ph]}}.

\bibitem{Giudice:2016zpa}
G.~F. Giudice, M.~McCullough, and A.~Urbano, ``{Hunting for Dark Particles with
  Gravitational Waves},''
  \href{http://dx.doi.org/10.1088/1475-7516/2016/10/001}{{\em JCAP} {\bfseries
  1610} no.~10, (2016) 001},
\href{http://arxiv.org/abs/1605.01209}{{\ttfamily arXiv:1605.01209 [hep-ph]}}.

\bibitem{Barack:2003fp}
L.~Barack and C.~Cutler, ``{LISA capture sources: Approximate waveforms,
  signal-to-noise ratios, and parameter estimation accuracy},''
  \href{http://dx.doi.org/10.1103/PhysRevD.69.082005}{{\em Phys. Rev.}
  {\bfseries D69} (2004) 082005},
\href{http://arxiv.org/abs/gr-qc/0310125}{{\ttfamily arXiv:gr-qc/0310125
  [gr-qc]}}.

\bibitem{Klein:2015hvg}
A.~Klein {\em et~al.}, ``{Science with the space-based interferometer eLISA:
  Supermassive black hole binaries},''
  \href{http://dx.doi.org/10.1103/PhysRevD.93.024003}{{\em Phys. Rev.}
  {\bfseries D93} no.~2, (2016) 024003},
\href{http://arxiv.org/abs/1511.05581}{{\ttfamily arXiv:1511.05581 [gr-qc]}}.

\bibitem{Ruffini:1969qy}
R.~Ruffini and S.~Bonazzola, ``{Systems of selfgravitating particles in general
  relativity and the concept of an equation of state},''
\href{http://dx.doi.org/10.1103/PhysRev.187.1767}{{\em Phys. Rev.} {\bfseries
  187} (1969) 1767--1783}.

\bibitem{Levkov:2016rkk}
D.~G. Levkov, A.~G. Panin, and I.~I. Tkachev, ``{Relativistic axions from
  collapsing Bose stars},''
  \href{http://dx.doi.org/10.1103/PhysRevLett.118.011301}{{\em Phys. Rev.
  Lett.} {\bfseries 118} no.~1, (2017) 011301},
\href{http://arxiv.org/abs/1609.03611}{{\ttfamily arXiv:1609.03611
  [astro-ph.CO]}}.

\bibitem{Gleiser:1988ih}
M.~Gleiser and R.~Watkins, ``{Gravitational Stability of Scalar Matter},''
\href{http://dx.doi.org/10.1016/0550-3213(89)90627-5}{{\em Nucl. Phys.}
  {\bfseries B319} (1989) 733--746}.

\bibitem{Gleiser:2009ys}
M.~Gleiser and D.~Sicilia, ``{A General Theory of Oscillon Dynamics},''
  \href{http://dx.doi.org/10.1103/PhysRevD.80.125037}{{\em Phys. Rev.}
  {\bfseries D80} (2009) 125037},
\href{http://arxiv.org/abs/0910.5922}{{\ttfamily arXiv:0910.5922 [hep-th]}}.

\bibitem{Eby:2015hyx}
J.~Eby, P.~Suranyi, and L.~C.~R. Wijewardhana, ``{The Lifetime of Axion
  Stars},'' \href{http://dx.doi.org/10.1142/S0217732316500905}{{\em Mod. Phys.
  Lett.} {\bfseries A31} no.~15, (2016) 1650090},
\href{http://arxiv.org/abs/1512.01709}{{\ttfamily arXiv:1512.01709 [hep-ph]}}.

\bibitem{Eby:2017teq}
J.~Eby, P.~Suranyi, and L.~C.~R. Wijewardhana, ``{Expansion in Higher Harmonics
  of Boson Stars using a Generalized Ruffini-Bonazzola Approach, Part 1: Bound
  States},'' \href{http://dx.doi.org/10.1088/1475-7516/2018/04/038}{{\em JCAP}
  {\bfseries 1804} no.~04, (2018) 038},
\href{http://arxiv.org/abs/1712.04941}{{\ttfamily arXiv:1712.04941 [hep-ph]}}.

\bibitem{Visinelli:2017ooc}
L.~Visinelli, S.~Baum, J.~Redondo, K.~Freese, and F.~Wilczek, ``{Dilute and
  dense axion stars},''
  \href{http://dx.doi.org/10.1016/j.physletb.2017.12.010}{{\em Phys. Lett.}
  {\bfseries B777} (2018) 64--72},
\href{http://arxiv.org/abs/1710.08910}{{\ttfamily arXiv:1710.08910
  [astro-ph.CO]}}.

\bibitem{Hertzberg:2018zte}
M.~P. Hertzberg and E.~D. Schiappacasse, ``{Dark Matter Axion Clump Resonance
  of Photons},'' \href{http://dx.doi.org/10.1088/1475-7516/2018/11/004}{{\em
  JCAP} {\bfseries 1811} no.~11, (2018) 004},
\href{http://arxiv.org/abs/1805.00430}{{\ttfamily arXiv:1805.00430 [hep-ph]}}.

\bibitem{Hawley:2000dt}
S.~H. Hawley and M.~W. Choptuik, ``{Boson stars driven to the brink of black
  hole formation},'' \href{http://dx.doi.org/10.1103/PhysRevD.62.104024}{{\em
  Phys. Rev.} {\bfseries D62} (2000) 104024},
\href{http://arxiv.org/abs/gr-qc/0007039}{{\ttfamily arXiv:gr-qc/0007039
  [gr-qc]}}.

\bibitem{Braaten:2015eeu}
E.~Braaten, A.~Mohapatra, and H.~Zhang, ``{Dense Axion Stars},''
  \href{http://dx.doi.org/10.1103/PhysRevLett.117.121801}{{\em Phys. Rev.
  Lett.} {\bfseries 117} no.~12, (2016) 121801},
\href{http://arxiv.org/abs/1512.00108}{{\ttfamily arXiv:1512.00108 [hep-ph]}}.

\bibitem{Schiappacasse:2017ham}
E.~D. Schiappacasse and M.~P. Hertzberg, ``{Analysis of Dark Matter Axion
  Clumps with Spherical Symmetry},''
  \href{http://dx.doi.org/10.1088/1475-7516/2018/03/E01,
  10.1088/1475-7516/2018/01/037}{{\em JCAP} {\bfseries 1801} (2018) 037},
  \href{http://arxiv.org/abs/1710.04729}{{\ttfamily arXiv:1710.04729
  [hep-ph]}}.
[Erratum: JCAP1803,no.03,E01(2018)].

\bibitem{Chavanis:2017loo}
P.-H. Chavanis, ``{Phase transitions between dilute and dense axion stars},''
  \href{http://dx.doi.org/10.1103/PhysRevD.98.023009}{{\em Phys. Rev.}
  {\bfseries D98} no.~2, (2018) 023009},
\href{http://arxiv.org/abs/1710.06268}{{\ttfamily arXiv:1710.06268 [gr-qc]}}.

\bibitem{Fan:2016rda}
J.~Fan, ``{Ultralight Repulsive Dark Matter and BEC},''
  \href{http://dx.doi.org/10.1016/j.dark.2016.10.005}{{\em Phys. Dark Univ.}
  {\bfseries 14} (2016) 84--94},
\href{http://arxiv.org/abs/1603.06580}{{\ttfamily arXiv:1603.06580 [hep-ph]}}.

\bibitem{Ferrarese:2000se}
L.~Ferrarese and D.~Merritt, ``{A Fundamental relation between supermassive
  black holes and their host galaxies},''
  \href{http://dx.doi.org/10.1086/312838}{{\em Astrophys. J.} {\bfseries 539}
  (2000) L9},
\href{http://arxiv.org/abs/astro-ph/0006053}{{\ttfamily arXiv:astro-ph/0006053
  [astro-ph]}}.

\bibitem{Gebhardt:2000fk}
K.~Gebhardt {\em et~al.}, ``{A Relationship between nuclear black hole mass and
  galaxy velocity dispersion},'' \href{http://dx.doi.org/10.1086/312840}{{\em
  Astrophys. J.} {\bfseries 539} (2000) L13},
\href{http://arxiv.org/abs/astro-ph/0006289}{{\ttfamily arXiv:astro-ph/0006289
  [astro-ph]}}.

\bibitem{Tremaine:2002js}
S.~Tremaine {\em et~al.}, ``{The slope of the black hole mass versus velocity
  dispersion correlation},'' \href{http://dx.doi.org/10.1086/341002}{{\em
  Astrophys. J.} {\bfseries 574} (2002) 740--753},
\href{http://arxiv.org/abs/astro-ph/0203468}{{\ttfamily arXiv:astro-ph/0203468
  [astro-ph]}}.

\bibitem{Alexander:2005jz}
T.~Alexander, ``{Stellar processes near the massive black hole in the Galactic
  Center},'' \href{http://dx.doi.org/10.1016/j.physrep.2005.08.002}{{\em Phys.
  Rept.} {\bfseries 419} (2005) 65--142},
\href{http://arxiv.org/abs/astro-ph/0508106}{{\ttfamily arXiv:astro-ph/0508106
  [astro-ph]}}.

\bibitem{alexander:review}
T.~Alexander, ``Stellar dynamics and stellar phenomena near a massive black
  hole,'' \href{http://dx.doi.org/10.1146/annurev-astro-091916-055306}{{\em
  Annual Review of Astronomy and Astrophysics} {\bfseries 55} no.~1, (2017)
  17--57},
  \href{http://arxiv.org/abs/https://doi.org/10.1146/annurev-astro-091916-055306}{{\ttfamily
  https://doi.org/10.1146/annurev-astro-091916-055306}}.
  \url{https://doi.org/10.1146/annurev-astro-091916-055306}.

\bibitem{Babak:2017tow}
S.~Babak, J.~Gair, A.~Sesana, E.~Barausse, C.~F. Sopuerta, C.~P.~L. Berry,
  E.~Berti, P.~Amaro-Seoane, A.~Petiteau, and A.~Klein, ``{Science with the
  space-based interferometer LISA. V: Extreme mass-ratio inspirals},''
\href{http://arxiv.org/abs/1703.09722}{{\ttfamily arXiv:1703.09722 [gr-qc]}}.

\bibitem{Babak:2006uv}
S.~Babak, H.~Fang, J.~R. Gair, K.~Glampedakis, and S.~A. Hughes, ``{'Kludge'
  gravitational waveforms for a test-body orbiting a Kerr black hole},''
  \href{http://dx.doi.org/10.1103/PhysRevD.75.024005,
  10.1103/PhysRevD.77.04990}{{\em Phys. Rev.} {\bfseries D75} (2007) 024005},
  \href{http://arxiv.org/abs/gr-qc/0607007}{{\ttfamily arXiv:gr-qc/0607007
  [gr-qc]}}.
[Erratum: Phys. Rev.D77,04990(2008)].

\bibitem{Chua:2017ujo}
A.~J.~K. Chua, C.~J. Moore, and J.~R. Gair, ``{Augmented kludge waveforms for
  detecting extreme-mass-ratio inspirals},''
  \href{http://dx.doi.org/10.1103/PhysRevD.96.044005}{{\em Phys. Rev.}
  {\bfseries D96} no.~4, (2017) 044005},
\href{http://arxiv.org/abs/1705.04259}{{\ttfamily arXiv:1705.04259 [gr-qc]}}.

\bibitem{Finn:2000sy}
L.~S. Finn and K.~S. Thorne, ``{Gravitational waves from a compact star in a
  circular, inspiral orbit, in the equatorial plane of a massive, spinning
  black hole, as observed by LISA},''
  \href{http://dx.doi.org/10.1103/PhysRevD.62.124021}{{\em Phys. Rev.}
  {\bfseries D62} (2000) 124021},
\href{http://arxiv.org/abs/gr-qc/0007074}{{\ttfamily arXiv:gr-qc/0007074
  [gr-qc]}}.

\bibitem{Teukolsky:1973ha}
S.~A. Teukolsky, ``{Perturbations of a rotating black hole. 1. Fundamental
  equations for gravitational electromagnetic and neutrino field
  perturbations},''
\href{http://dx.doi.org/10.1086/152444}{{\em Astrophys. J.} {\bfseries 185}
  (1973) 635--647}.

\bibitem{Sasaki:1981sx}
M.~Sasaki and T.~Nakamura, ``{Gravitational Radiation From a Kerr Black Hole.
  1. Formulation and a Method for Numerical Analysis},''
\href{http://dx.doi.org/10.1143/PTP.67.1788}{{\em Prog. Theor. Phys.}
  {\bfseries 67} (1982) 1788}.

\bibitem{Randall:2017jop}
L.~Randall and Z.-Z. Xianyu, ``{Induced Ellipticity for Inspiraling Binary
  Systems},'' \href{http://dx.doi.org/10.3847/1538-4357/aaa1a2}{{\em Astrophys.
  J.} {\bfseries 853} no.~1, (2018) 93},
\href{http://arxiv.org/abs/1708.08569}{{\ttfamily arXiv:1708.08569 [gr-qc]}}.

\bibitem{Moore:2014lga}
C.~J. Moore, R.~H. Cole, and C.~P.~L. Berry, ``{Gravitational-wave sensitivity
  curves},'' \href{http://dx.doi.org/10.1088/0264-9381/32/1/015014}{{\em Class.
  Quant. Grav.} {\bfseries 32} no.~1, (2015) 015014},
\href{http://arxiv.org/abs/1408.0740}{{\ttfamily arXiv:1408.0740 [gr-qc]}}.

\bibitem{Babak:2009cj}
{\bfseries Mock LISA Data Challenge Task Force} Collaboration, S.~Babak {\em
  et~al.}, ``{The Mock LISA Data Challenges: From Challenge 3 to Challenge
  4},'' \href{http://dx.doi.org/10.1088/0264-9381/27/8/084009}{{\em Class.
  Quant. Grav.} {\bfseries 27} (2010) 084009},
\href{http://arxiv.org/abs/0912.0548}{{\ttfamily arXiv:0912.0548 [gr-qc]}}.

\bibitem{Bardeen:1972fi}
J.~M. Bardeen, W.~H. Press, and S.~A. Teukolsky, ``{Rotating black holes:
  Locally nonrotating frames, energy extraction, and scalar synchrotron
  radiation},''
\href{http://dx.doi.org/10.1086/151796}{{\em Astrophys. J.} {\bfseries 178}
  (1972) 347}.

\bibitem{Gong:2014mca}
X.~Gong {\em et~al.}, ``{Descope of the ALIA mission},''
  \href{http://dx.doi.org/10.1088/1742-6596/610/1/012011}{{\em J. Phys. Conf.
  Ser.} {\bfseries 610} no.~1, (2015) 012011},
\href{http://arxiv.org/abs/1410.7296}{{\ttfamily arXiv:1410.7296 [gr-qc]}}.

\bibitem{Luo:2015ght}
{\bfseries TianQin} Collaboration, J.~Luo {\em et~al.}, ``{TianQin: a
  space-borne gravitational wave detector},''
  \href{http://dx.doi.org/10.1088/0264-9381/33/3/035010}{{\em Class. Quant.
  Grav.} {\bfseries 33} no.~3, (2016) 035010},
\href{http://arxiv.org/abs/1512.02076}{{\ttfamily arXiv:1512.02076
  [astro-ph.IM]}}.

\bibitem{Kudoh:2005as}
H.~Kudoh, A.~Taruya, T.~Hiramatsu, and Y.~Himemoto, ``{Detecting a
  gravitational-wave background with next-generation space interferometers},''
  \href{http://dx.doi.org/10.1103/PhysRevD.73.064006}{{\em Phys. Rev.}
  {\bfseries D73} (2006) 064006},
\href{http://arxiv.org/abs/gr-qc/0511145}{{\ttfamily arXiv:gr-qc/0511145
  [gr-qc]}}.

\bibitem{Dominik:2014yma}
M.~Dominik, E.~Berti, R.~O'Shaughnessy, I.~Mandel, K.~Belczynski, C.~Fryer,
  D.~E. Holz, T.~Bulik, and F.~Pannarale, ``{Double Compact Objects III:
  Gravitational Wave Detection Rates},''
  \href{http://dx.doi.org/10.1088/0004-637X/806/2/263}{{\em Astrophys. J.}
  {\bfseries 806} no.~2, (2015) 263},
\href{http://arxiv.org/abs/1405.7016}{{\ttfamily arXiv:1405.7016
  [astro-ph.HE]}}.

\bibitem{stone:2015tidal}
N.~C. Stone, ``{The Tidal Disruption of Stars by Supermassive Black Holes},''
  \href{http://dx.doi.org/DOI 10.1007/978-3-319-12676-0}{{\em Springer Theses}
  (2015) 1--162}.

\bibitem{Damour:2009vw}
T.~Damour and A.~Nagar, ``{Relativistic tidal properties of neutron stars},''
  \href{http://dx.doi.org/10.1103/PhysRevD.80.084035}{{\em Phys. Rev.}
  {\bfseries D80} (2009) 084035},
\href{http://arxiv.org/abs/0906.0096}{{\ttfamily arXiv:0906.0096 [gr-qc]}}.

\bibitem{Cardoso:2017cfl}
V.~Cardoso, E.~Franzin, A.~Maselli, P.~Pani, and G.~Raposo, ``{Testing
  strong-field gravity with tidal Love numbers},''
  \href{http://dx.doi.org/10.1103/PhysRevD.95.089901,
  10.1103/PhysRevD.95.084014}{{\em Phys. Rev.} {\bfseries D95} no.~8, (2017)
  084014}, \href{http://arxiv.org/abs/1701.01116}{{\ttfamily arXiv:1701.01116
  [gr-qc]}}.
[Addendum: Phys. Rev.D95,no.8,089901(2017)].

\bibitem{Binnington:2009bb}
T.~Binnington and E.~Poisson, ``{Relativistic theory of tidal Love numbers},''
  \href{http://dx.doi.org/10.1103/PhysRevD.80.084018}{{\em Phys. Rev.}
  {\bfseries D80} (2009) 084018},
\href{http://arxiv.org/abs/0906.1366}{{\ttfamily arXiv:0906.1366 [gr-qc]}}.

\bibitem{Mielke:2016war}
E.~W. Mielke, ``{Rotating Boson Stars},''
\href{http://dx.doi.org/10.1007/978-3-319-31299-6_6}{{\em Fundam. Theor. Phys.}
  {\bfseries 183} (2016) 115--131}.

\bibitem{Sennett:2017etc}
N.~Sennett, T.~Hinderer, J.~Steinhoff, A.~Buonanno, and S.~Ossokine,
  ``{Distinguishing Boson Stars from Black Holes and Neutron Stars from Tidal
  Interactions in Inspiraling Binary Systems},''
  \href{http://dx.doi.org/10.1103/PhysRevD.96.024002}{{\em Phys. Rev.}
  {\bfseries D96} no.~2, (2017) 024002},
\href{http://arxiv.org/abs/1704.08651}{{\ttfamily arXiv:1704.08651 [gr-qc]}}.

\bibitem{Komossa:2015qya}
S.~Komossa, ``{Tidal disruption of stars by supermassive black holes: Status of
  observations},'' \href{http://dx.doi.org/10.1016/j.jheap.2015.04.006}{{\em
  JHEAp} {\bfseries 7} (2015) 148--157},
\href{http://arxiv.org/abs/1505.01093}{{\ttfamily arXiv:1505.01093
  [astro-ph.HE]}}.

\bibitem{Lodato:2015aoa}
G.~Lodato, A.~Franchini, C.~Bonnerot, and E.~M. Rossi, ``{Recent developments
  in the theory of tidal disruption events},''
\href{http://dx.doi.org/10.1016/j.jheap.2015.04.003}{{\em JHEAp} {\bfseries 7}
  (2015) 158--162}.

\bibitem{Navarro:1996bv}
J.~F. Navarro, V.~R. Eke, and C.~S. Frenk, ``{The cores of dwarf galaxy
  halos},'' \href{http://dx.doi.org/10.1093/mnras/283.3.72L,
  10.1093/mnras/283.3.L72}{{\em Mon. Not. Roy. Astron. Soc.} {\bfseries 283}
  (1996) L72--L78},
\href{http://arxiv.org/abs/astro-ph/9610187}{{\ttfamily arXiv:astro-ph/9610187
  [astro-ph]}}.

\bibitem{Pontzen:2011ty}
A.~Pontzen and F.~Governato, ``{How supernova feedback turns dark matter cusps
  into cores},'' \href{http://dx.doi.org/10.1111/j.1365-2966.2012.20571.x}{{\em
  Mon. Not. Roy. Astron. Soc.} {\bfseries 421} (2012) 3464},
\href{http://arxiv.org/abs/1106.0499}{{\ttfamily arXiv:1106.0499
  [astro-ph.CO]}}.

\bibitem{Tulin:2017ara}
S.~Tulin and H.-B. Yu, ``{Dark Matter Self-interactions and Small Scale
  Structure},''
\href{http://arxiv.org/abs/1705.02358}{{\ttfamily arXiv:1705.02358 [hep-ph]}}.

\bibitem{Schive:2014hza}
H.-Y. Schive, M.-H. Liao, T.-P. Woo, S.-K. Wong, T.~Chiueh, T.~Broadhurst, and
  W.~Y.~P. Hwang, ``{Understanding the Core-Halo Relation of Quantum Wave Dark
  Matter from 3D Simulations},''
  \href{http://dx.doi.org/10.1103/PhysRevLett.113.261302}{{\em Phys. Rev.
  Lett.} {\bfseries 113} no.~26, (2014) 261302},
\href{http://arxiv.org/abs/1407.7762}{{\ttfamily arXiv:1407.7762
  [astro-ph.GA]}}.

\bibitem{Hui:2016ltb}
L.~Hui, J.~P. Ostriker, S.~Tremaine, and E.~Witten, ``{Ultralight scalars as
  cosmological dark matter},''
  \href{http://dx.doi.org/10.1103/PhysRevD.95.043541}{{\em Phys. Rev.}
  {\bfseries D95} no.~4, (2017) 043541},
\href{http://arxiv.org/abs/1610.08297}{{\ttfamily arXiv:1610.08297
  [astro-ph.CO]}}.

\bibitem{Navarro:1995iw}
J.~F. Navarro, C.~S. Frenk, and S.~D.~M. White, ``{The Structure of cold dark
  matter halos},'' \href{http://dx.doi.org/10.1086/177173}{{\em Astrophys. J.}
  {\bfseries 462} (1996) 563--575},
\href{http://arxiv.org/abs/astro-ph/9508025}{{\ttfamily arXiv:astro-ph/9508025
  [astro-ph]}}.

\bibitem{Navarro:1996gj}
J.~F. Navarro, C.~S. Frenk, and S.~D.~M. White, ``{A Universal density profile
  from hierarchical clustering},'' \href{http://dx.doi.org/10.1086/304888}{{\em
  Astrophys. J.} {\bfseries 490} (1997) 493--508},
\href{http://arxiv.org/abs/astro-ph/9611107}{{\ttfamily arXiv:astro-ph/9611107
  [astro-ph]}}.

\bibitem{Gondolo:1999ef}
P.~Gondolo and J.~Silk, ``{Dark matter annihilation at the galactic center},''
  \href{http://dx.doi.org/10.1103/PhysRevLett.83.1719}{{\em Phys. Rev. Lett.}
  {\bfseries 83} (1999) 1719--1722},
\href{http://arxiv.org/abs/astro-ph/9906391}{{\ttfamily arXiv:astro-ph/9906391
  [astro-ph]}}.

\bibitem{Sadeghian:2013laa}
L.~Sadeghian, F.~Ferrer, and C.~M. Will, ``{Dark matter distributions around
  massive black holes: A general relativistic analysis},''
  \href{http://dx.doi.org/10.1103/PhysRevD.88.063522}{{\em Phys. Rev.}
  {\bfseries D88} no.~6, (2013) 063522},
\href{http://arxiv.org/abs/1305.2619}{{\ttfamily arXiv:1305.2619
  [astro-ph.GA]}}.

\bibitem{Sandick:2016zeg}
P.~Sandick, K.~Sinha, and T.~Yamamoto, ``{Black Holes, Dark Matter Spikes, and
  Constraints on Simplified Models with $t$-Channel Mediators},''
  \href{http://dx.doi.org/10.1103/PhysRevD.98.035004}{{\em Phys. Rev.}
  {\bfseries D98} no.~3, (2018) 035004},
\href{http://arxiv.org/abs/1701.00067}{{\ttfamily arXiv:1701.00067 [hep-ph]}}.

\bibitem{Ferrer:2017xwm}
F.~Ferrer, A.~M. da~Rosa, and C.~M. Will, ``{Dark matter spikes in the vicinity
  of Kerr black holes},''
\href{http://arxiv.org/abs/1707.06302}{{\ttfamily arXiv:1707.06302
  [astro-ph.CO]}}.

\bibitem{gahcall:1976aa}
J.~N. Bahcall and R.~A. Wolf, ``{Star distribution around a massive black hole
  in a globular cluster},''
\href{http://dx.doi.org/10.1086/154711}{{\em Astrophys. J.} {\bfseries 209}
  (1976) 214--232}.

\bibitem{Hopman:2006xn}
C.~Hopman and T.~Alexander, ``{The effect of mass-segregation on gravitational
  wave sources near massive black holes},''
  \href{http://dx.doi.org/10.1086/506273}{{\em Astrophys. J.} {\bfseries 645}
  (2006) L133--L136},
\href{http://arxiv.org/abs/astro-ph/0603324}{{\ttfamily arXiv:astro-ph/0603324
  [astro-ph]}}.

\bibitem{AmaroSeoane:2010bq}
P.~Amaro-Seoane and M.~Preto, ``{The impact of realistic models of mass
  segregation on the event rate of extreme-mass ratio inspirals and cusp
  re-growth},'' \href{http://dx.doi.org/10.1088/0264-9381/28/9/094017}{{\em
  Class. Quant. Grav.} {\bfseries 28} (2011) 094017},
\href{http://arxiv.org/abs/1010.5781}{{\ttfamily arXiv:1010.5781
  [astro-ph.CO]}}.

\bibitem{Alexander:2008tq}
T.~Alexander and C.~Hopman, ``{Strong mass segregation around a massive black
  hole},'' \href{http://dx.doi.org/10.1088/0004-637X/697/2/1861}{{\em
  Astrophys. J.} {\bfseries 697} (2009) 1861--1869},
\href{http://arxiv.org/abs/0808.3150}{{\ttfamily arXiv:0808.3150 [astro-ph]}}.

\bibitem{Guo:2017njn}
H.-K. Guo, J.~Shu, and Y.~Zhao, ``{Using LISA-like Gravitational Wave Detectors
  to Search for Primordial Black Holes},''
  \href{http://dx.doi.org/10.1103/PhysRevD.99.023001}{{\em Phys. Rev.}
  {\bfseries D99} no.~2, (2019) 023001},
\href{http://arxiv.org/abs/1709.03500}{{\ttfamily arXiv:1709.03500
  [astro-ph.CO]}}.

\end{thebibliography}\endgroup

\end{document}